 \documentclass[3p,times,twocolumn,authoryear]{elsarticle}

\usepackage{amssymb,ulem,xcolor,url}
\usepackage{arydshln}
\usepackage{amsmath}

\usepackage[switch]{lineno}
\usepackage{lscape}
\usepackage{overpic}

\journal{Journal of High Energy Astrophysics}

\begin{document}

\begin{frontmatter}



\title{Exploring the supernova remnant contribution to the first LHAASO source catalog via passively illuminated interstellar clouds} 


\author[ecap]{Alison M.W. Mitchell} 
\ead{alison.mw.mitchell@fau.de}
\author[sapienza]{Silvia Celli}
\ead{silvia.celli@uniroma1.it}

\affiliation[ecap]{organization={Erlangen Centre for Astroparticle Physics, Friedrich-Alexander-University Erlangen-Nuremberg}, 
            addressline={Nikolaus-Fiebiger-Str. 2}, 
            city={Erlangen},
            postcode={91058}, 
            country={Germany}}

\affiliation[sapienza]{organization={Physics Department, Sapienza University of Rome},
            addressline={Piazzale Aldo Moro 2}, 
            city={Rome},
            postcode={00185}, 
            country={Italy}}

\begin{abstract}
Supernova remnants (SNRs) are considered as the most promising source class to account for the bulk of the Galactic cosmic-ray flux. Yet amongst the population of ultra-high energy (UHE) sources that has recently emerged, due to high-altitude particle detector experiments such as LHAASO and HAWC, remarkably few are associated with known SNRs. These observations might well indicate that the highest energy particles would escape the remnant early during the shock evolution as a result of its reduced confinement capabilities. This flux of escaping particles may then encounter dense targets (gas and dust) for hadronic interactions in the form of both atomic and molecular material such as interstellar clouds, thereby generating a UHE gamma-ray flux. We explore such a scenario here, considering known SNRs in a physically driven model for particle escape, and as coupled to molecular clouds in the Galaxy. Our analysis allows the investigation of SNR-illuminated clouds in coincidence with sources detected in the first LHAASO catalogue. Indeed, the illuminated interstellar clouds may contribute to the total gamma-ray flux from several unidentified sources, as we discuss here. Yet we nevertheless find that further detailed studies will be necessary to verify or refute this scenario of passive UHE gamma-ray sources in future. 
\end{abstract}


\end{frontmatter}


\section{Introduction}
\label{sec:intro}

Considerable experimental advancements in recent years have led to the emergence of UHE ($E>100$\,TeV) gamma-ray astronomy, most notably due to ground-based particle detectors such as Tibet-AS$\gamma$ \citep{2009ApJ...692...61A_tibetAScrab}, the HAWC (High Altitude Water Cherenkov) observatory \citep{2017ApJ...843...39A_hawcCrab}, and LHAASO \citep{Aharonian_2021_lhaasoCrab}. UHE gamma rays are of particular interest to investigate the origins of Cosmic Rays (CRs) as the production of energetic photons is necessarily due to primaries with PeV energies. Galactic sources of UHE gamma rays therefore provide strong indications as to the locations of Galactic PeVatrons, nearby accelerators of CRs to the $\sim3\times 10^{15}$\,eV energy of the `knee' spectral break, which are thought to be the most powerful accelerators in our Galaxy. 
Energetic arguments indicate that supernovae (SNe) can provide the bulk of the CR flux, however, there remain theoretical difficulties with achieving PeV energies via diffuse shock acceleration in SNRs \citep{1978MNRAS.182..147Bell}. Nevertheless, the considerable number of SNRs detected at gamma-ray energies, coupled to the evidence for the pion-bump spectral signature in some of these \citep{2010ApJ...710L.151T_Pion_IC443_Agile,2011ApJ...742L..30G_Pion_W44_Agile}, suggests that SNe indeed contribute to the CR flux considerably \citep{2018A&A...612A...3HESS_snrpop,2013Sci...339..807A_FermiPion}. What remains yet to be understood is up to which energy they constitute the dominant source population of the CR flux.

Although there's still a lack of direct and conclusive evidence of SNRs as hadronic PeVatrons in VHE and UHE observations (even with LHAASO), SNRs  still remain candidate PeV accelerators, as they  might be producing particles at PeV energies in short time intervals, namely during their early evolution. In particular, the self-amplification of the magnetic field by the current of escaping particles might provide the level of turbulence required to enhance the particle confinement, thus maintaining them in the acceleration region for longer than would otherwise be the case. However, as the shock speed decreases, magnetic turbulence drops such that the accelerated particles would not remain trapped to their accelerator, failing to provide detectable UHE gamma-ray flux at the present day \citep{2021Univ....7..324Cristofari}.

In this context, it was certainly surprising that the most prominent source class at UHEs is emerging to be the environments of energetic pulsars, typically pulsar wind nebulae (PWNe) \citep{2020PhRvL.124b1102A_HAWC56,2021Natur.594...33CaoUHE}. These observations provide a challenge to the leptonic origin of the emission via inverse Compton scattering, one that can nevertheless be alleviated in the case of intense radiation environments \citep{2021ApJ...908L..49Breuhaus,2022A&A...660A...8Breuhaus_lhaaso}. Depending on the PWN age, the associated SNR might still be located in the very same region, calling for gamma-ray measurements with improved angular resolution to disentangle the different components \citep{celli_peron} and clearly unveil whether PWNe are more likely UHE emitters than SNRs, as putatively indicated by LHAASO observations. At the same time, one should recall that young and massive star clusters have recently been shown to be suitable CR factories at PeV energies, such that their overall contribution to Galactic CRs remains to be clarified \citep{2021MNRAS.504.6096MorlinoSC}. The limited amount of observations at very-high and UHEs so far available for this source population are consistent with expectations derived for local open clusters \citep{gaiaSC}, requiring next generation imaging instruments to enlarge the statistics. \\

The majority of UHE emitters are, however, currently unidentified (UNID), in that the cause of the emission is not known or ambiguous.
Additionally, a number of UHE emitters are unassociated or `dark' in origin, referring to the lack of known counterparts. A noticeable interest resides in these dark systems, as their emission might arise as a result of the illumination by nearby accelerators through a particle beam colliding in there. For example, \cite{GabiciAharonianPeVatron07} posited that gas overdensities (such as molecular clouds) in the vicinity of powerful accelerators can act as a target for particle interactions. 
To explore the possibility that the dark systems are indeed passive targets of SNRs, we proceed by investigating physical associations between the known population of SNRs and molecular clouds, consisting of both spatial and spectral correlation. The former strongly relies on the catalogued positions of both SNRs and clouds, which are provided as 3D information. On the other hand, a spectral connection among two class members can be established based on a physical model describing acceleration in SNRs, transport to the cloud location, and hadronic collisions in the target cloud.
As the most energetic particles will escape from the accelerator at early times, rather than remaining at source, spectral signatures from these can be found associated with interstellar clouds. Indeed, using such clouds to search for PeVatrons is a viable method to unveil Galactic CR accelerators, that we started exploring in \cite{2021MNRAS.503.3522Mitchell} and complement here. In principle, one may even anticipate a new population of such sources emerging at the highest energies, without GeV-TeV counterparts. 

For the specific case of the dark UHE source LHAASO\,J2108+5157 the scenario of a molecular cloud being illuminated by an as yet unknown accelerator (e.g. an SNR) was found to be compatible with the detected emission \citep{2021ApJ...919L..22Cao_J2108,2024A&A...684A..66Mitchell}. 

Although interstellar clouds may also shine in gamma rays due to interactions from the sea of Galactic CRs \citep{peron}, it is unlikely that this contribution can account for any of the UHE sources, due to the significant level of CR flux enhancement that would be necessary at $E\gtrsim 10^{14}$\,eV, as clearly demonstrated in \citet{celli_peron}.

In this study, we investigate if further UHE and/or unidentified sources from \cite{2024ApJS..271...25Cao_1lhaaso} can be explained as molecular clouds illuminated by nearby SNRs, based on the results of \cite{2021MNRAS.503.3522Mitchell}. 
This paper is organised as follows. Section \ref{sec:method} briefly describes the SNR-cloud datasets and the model used for particle acceleration, propagation and interaction in clouds to predict the expected gamma-ray flux of these systems. In section \ref{sec:population} we compare our results with sources from the first LHAASO catalogue (1LHAASO) \citep{2024ApJS..271...25Cao_1lhaaso}, searching for spatial correlations. Section \ref{sec:sources} looks in more detail at  1LHAASO unidentified sources, comparing their spectra with the expected emission from overlapping clouds, in the scenario that these are illuminated by nearby SNRs. We discuss the interpretation and model uncertainties, before concluding in section \ref{sec:conclude}.

\begin{figure*}
    \centering
    \includegraphics[width=2\columnwidth]{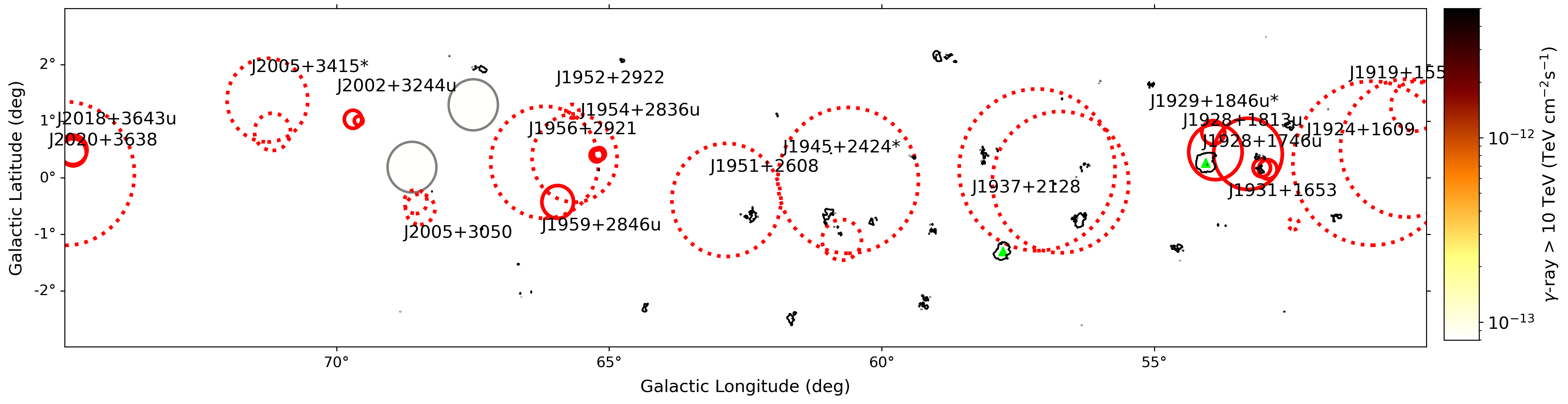}
    \includegraphics[width=2\columnwidth]{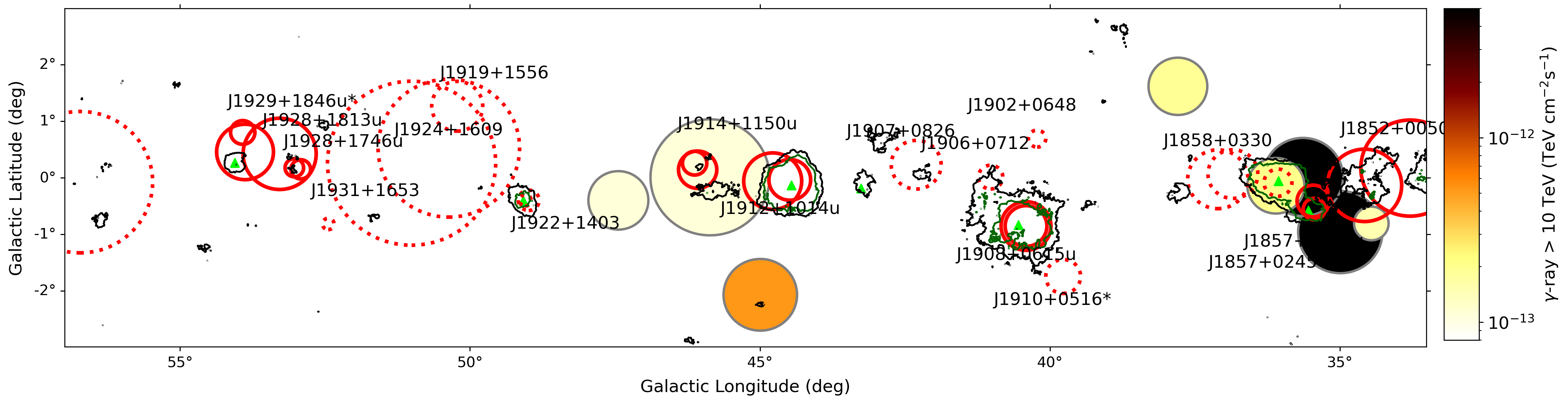}
    \includegraphics[width=2\columnwidth]{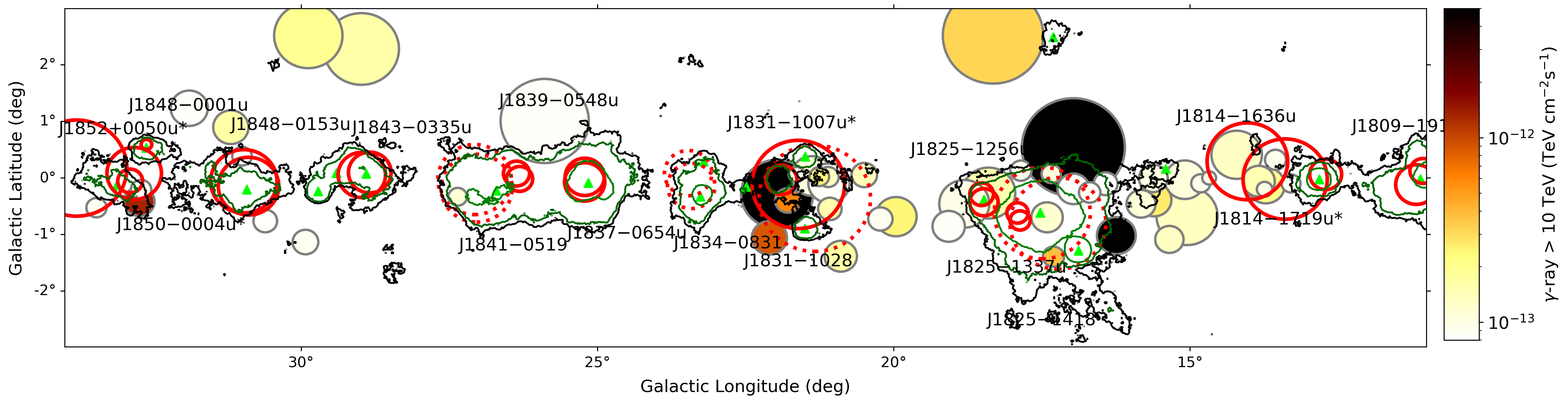}
    \caption{The predicted integral gamma-ray flux above 10\,TeV from molecular clouds of the \citet{2016ApJ...822...52Rice}, due to illumination by nearby SNRs (all assumed to be of type II). Significance contours (green and black) from the HGPS are reported together with name and locations of 1LHAASO sources (red circles), with solid red circles indicating UHE sources (those with high-significance emission reaching beyond $\sim100$\,TeV). }
    \label{fig:galplane}
\end{figure*}

\begin{figure*}
    \centering
    \includegraphics[width=2\columnwidth]{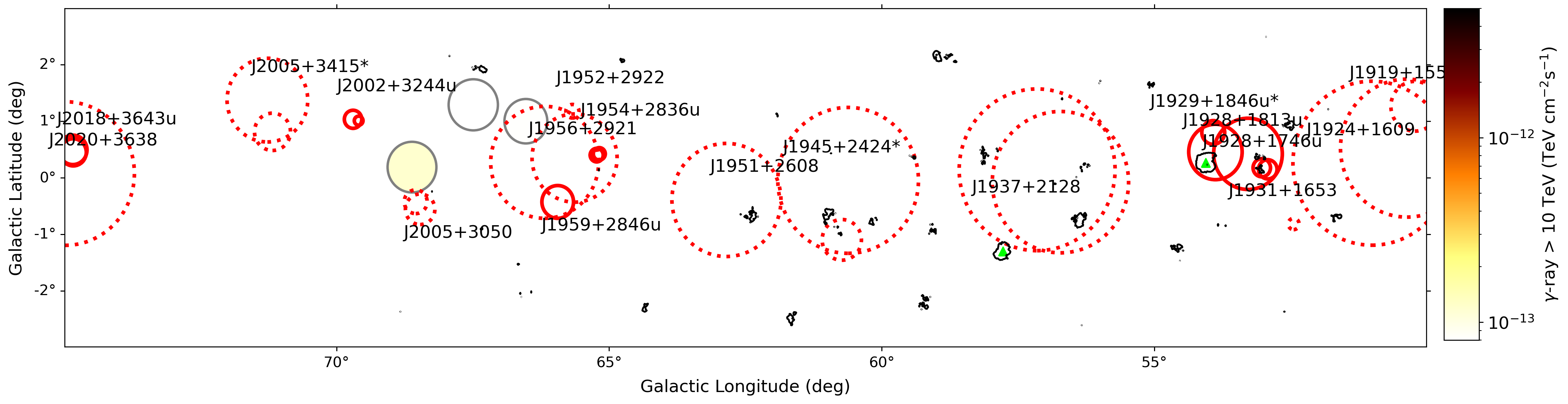}
    \includegraphics[width=2\columnwidth]{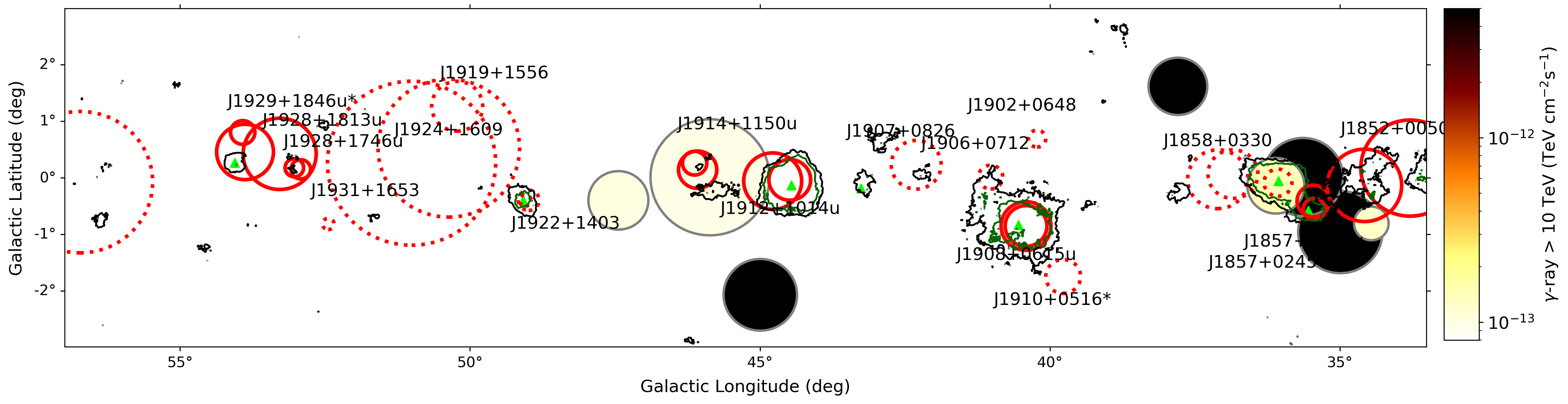}
    \includegraphics[width=2\columnwidth]{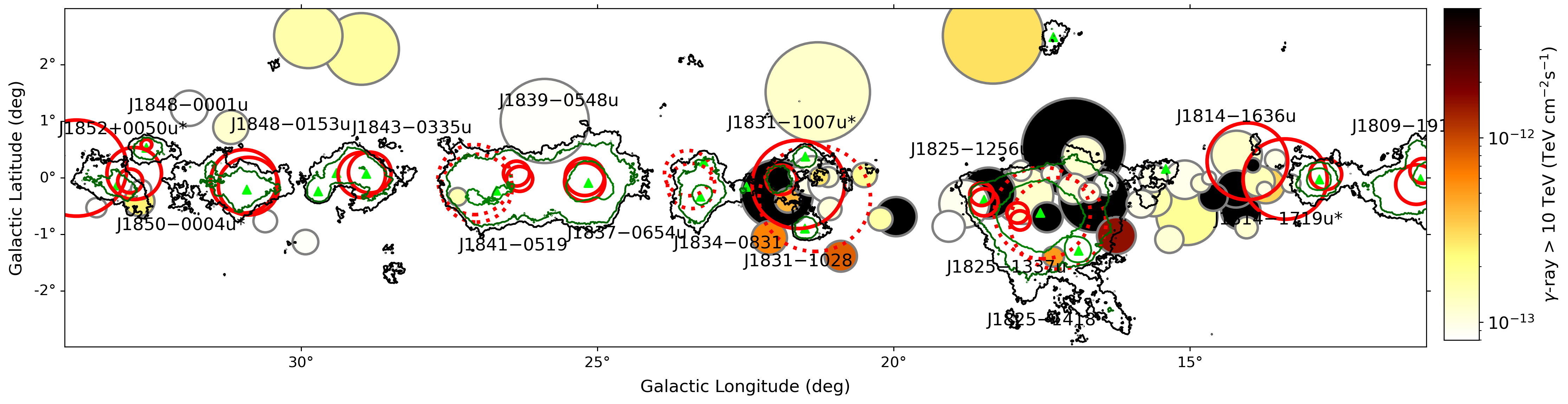}
    \caption{As in figure \ref{fig:galplane} with color scale referring to the model of particle escape from type Ia SNe.}
    \label{fig:galplaneIa}
\end{figure*}

\section{Methodology}
\label{sec:method}

The model for impulsive CR acceleration in SNRs, transport through the intervening interstellar medium (ISM) and interaction with nearby clouds is described in detail in \cite{2021MNRAS.503.3522Mitchell}. It is based on the proton injection model of \cite{AA96} and generation of gamma-ray flux as described by \cite{Kelner06}. In particular, we adopt a proton acceleration slope equal to 2.0 for the differential energy spectrum and consider an energy-dependent particle escape commencing at the Sedov time, $t_{\rm sed}$, marking the start of the adiabatic SNR expansion. We assume that the SNR radius expands with time as $R_{\rm SNR}\propto t^{2/5}$ during the Sedov-Taylor phase \citep{sedov1959}. 
We use two catalogues of SNRs, Green's catalogue \citep{2019JApA...40...36Green} and the SNRcat \citep{snrcat} to obtain the corresponding properties (location, age, distance and physical size) required as inputs to our model.  
For Galactic molecular clouds, we adopt the \cite{2016ApJ...822...52Rice} catalogue, providing size and 3D position of CO over-densities in the plane of our Galaxy. Typically, these are $\sim 10-100\,{\rm cm}^{-3}$ with radii on the order of tens of parsecs. We note that the catalogue of \cite{2017ApJ...834...57Miville} is based on the same underlying CO dataset \citep{Dame01}, but includes a large number of smaller clouds. Therefore, we use \cite{2016ApJ...822...52Rice} to focus on the most massive clouds in the first instance for this study. 

For cases in which no estimate of the SNR age is provided in either catalogues, we use the measured size of the SNR to obtain an estimation of its age. For cases in which the distance to the SNR is unknown, for the purposes of our model we assume that the SNR is located at the same distance from Earth as the molecular cloud with which it has been spatially paired using the sky coordinates. 

Our descriptions of particle acceleration in remnants of type II and type Ia SNe mainly differs in the onset of the Sedov stage: fixing the energetics and upstream density, assumed to be uniform in both cases, the Sedov time is shorter for thermonuclear explosions because of the reduced ejecta mass. For a $10^{51}$~erg explosion in an ISM numerical density of $n_{\rm ISM}=1{\rm cm}^{-3}$ with a CR conversion efficiency of 10\%, we obtain $t_{\rm sed} \simeq 1.6$\,kyr in a type II SN with $10 \, M_\odot$ of ejecta, and $t_{\rm sed}\approx 234\,{\rm yr}$ for a type Ia with a $1 \, M_\odot$ ejecta. Despite the fact that the assumption of uniformity of the surrounding ISM might not formally hold in the case of type II explosions, we proceed here with this simplification remarking that the Sedov time can actually be smaller than that estimated above once the correct density profile excavated by the progenitor's wind is accounted for \citep{2015APh....69....1Cardillo}. Particle escape from the SNR shock front is time and momentum dependent, commencing at $t_{\rm sed}$ for the highest energy particles accelerated in the system \citep{gabici2009}. The behavior of particle escape can be described by a power-law dependence of the escape time $t_{\rm esc}$ with respect to the particle momentum $p$ as \citep{2019MNRAS.490.4317Celli}:
\begin{equation}
    t_{\rm esc}(p) = t_{\rm Sed} (p/p_M)^{-1/\beta}\,
\end{equation}
where the parameter $\beta$ embeds information about the evolution of the magnetic turbulence and its confinement properties. In this work, we set $\beta=2.5$ and $p_M=3$\,PeV/c, corresponding to the maximum energy reached at the Sedov time as required to achieve the CR knee. However, it is worth keeping in mind that both these parameters are unknown, and should be constrained individually for each SNR by means of spectral observations. For example, the maximum energy obtained in the system depends on the specificity of the explosion (shock speed and circumstellar density), and might well fall short of the value adoped here in many of the systems. Nonetheless, we adopt 3\,PeV as a reference to explore the PeVatron scenario.
The magnetic field strength is assumed to be 3\,$\mu$G in the ISM and 10\,$\mu$G within clouds, with measurements of the Zeeman effect in high density environments ($n \geq 300\,{\rm cm}^{-3}$) indicating enhancements as $B(n)\propto 10 \left(n/300\right)^{0.65}\,\mu{\rm G}$ \citep{2010ApJ...725..466Crutcher}.

The diffusive transport of particles through the ISM is parameterised through a diffusion coefficient as: 
\begin{equation}
    D(E) = \chi D_0 \left(\frac{E/{\rm GeV}}{B(n)/3\mu{\rm G}}\right)^\delta\,,
\end{equation}
where $D_0$ represents a reference value of diffusion coefficient at 1~GeV in the vicinity of the accelerator, such that $\chi=1$ in standard ISM. 
Due to the effects of CR-induced turbulence and self-confinement in the vicinity of Galactic accelerators, we adopted a normalisation of the ISM diffusion coefficient that is mildly suppressed with respect to the ISM average, namely $D_0=3\times10^{26}{\rm cm^{-2}s^{-1}}$ at 1\,GeV.
In turn, within interstellar clouds, the transport is treated as suppressed due to enhanced turbulence, with $\chi$ in the range $0.05-0.1$. For the purposes of the baseline scenario and for figures \ref{fig:galplane} and \ref{fig:galplaneIa}, we set $\chi=0.1$. We adopt an energy dependence of the diffusion coefficient of $\delta=0.5$, which is expected to result for Kraichnan-like cascading of turbulence and it appears to be consistent with local CR data. Note, however, that the exact domain of diffusion operating in the source vicinity, particularly in clouds, is an unknown property, such that realistic values of this parameter range within 0.3 (Kolmogorov-like) and 1 (Bohm-like). This constitutes an additional source of uncertainty in the model. 

The space, time and energy-dependent $\gamma$-ray emissivity $\Phi_\gamma$ resulting from interactions of an incident proton spectrum $f(E,r',t')$ with target material of density $n$ is described by:
\begin{equation}
    \Phi_\gamma  = cn\int^\infty_{E_\gamma} \sigma_{\rm inel}(E)f(E,r',t')F_\gamma\left(\frac{E_\gamma}{E},E\right)\frac{\mathrm{d}E}{E}
    \label{eq:phig}
\end{equation}
where $\sigma_{\rm inel}(E)$ is the energy-dependent inelastic cross-section for proton-proton interactions, that we solely consider in this work, and the kernel function $F_\gamma$ is taken from \citet{Kelner06}. The gamma-ray flux $F$ at Earth emerging from a cloud of volume $V_c$ located at a distance $d$ from the particle accelerator can be evaluated as:
\begin{equation}
    F(E_\gamma , t) = \Phi_\gamma(E_\gamma, t) V_c/ (4\pi d^2)\,.
\end{equation}

In addition to the particle flux $f(E,r',t')$ due to a nearby accelerator, we also consider the Galactic diffuse CR flux $f_G(E)$, simply summing the total particle contribution, i.e. by replacing $f(E,r',t')$ by $f(E,r',t')+f_G(E)$ in equation \eqref{eq:phig} \citep{PhysRevLett.114.171103_AMS}. Due to the steepness of the CR spectrum, the contribution of the Galactic diffuse CRs is, however, negligible at UHE with respect to that from a nearby accelerator.


\begin{table*}
\small
    \centering
    \begin{tabular}{c|cccccc|llcc}
    Cloud & ($l$,$b$) & R$_{cloud}$ & $n$ & D & $\phi_\gamma>10\,{\rm TeV}$ & SNR & 1LHAASO & inst. & R$_{39}$ & Sep. \\
     & ($^\circ$,$^\circ$) & ($^\circ$) & (cm$^{-3}$) & (kpc) & (TeV\,cm$^{-2}$s$^{-1}$) & & & & ($^\circ$) & ($^\circ$) \\
    \hline
    151 & (16.97, 0.53) & 0.46 & 90 & 2.08 & 2.76e-11 & G019.1+00.2 & J1825-1418 & WCDA & 0.81 & 1.37 \\
& & & & & & & & KM2A & 0.81 & 1.27 \\
    190 & (34.99, -0.96) & 0.38 & 126 & 2.62 & 1.76e-11 & G036.6-00.7 & J1852+0050u* & KM2A* & 0.85 & 1.65 \\
 & & & & & & &  & WCDA & 0.64 & 0.93 \\
 & & & & & & & {\bf J1857+0203u} & KM2A & 0.28 & 0.72 \\
 & & & & & & &  & WCDA & 0.19 & 0.61 \\
166 & (17.30, -1.40) & 0.09 & 652 & 3.14 & 7.12e-12 & G016.2-02.7 & J1825-1418 & WCDA & 0.81 & 0.59 \\
 & & & & & 1.1e-11 & G016.4-00.5 & & KM2A & 0.81 & 0.79 \\
 & & & & & 1.04e-12 & G017.4-02.3 &  & & & \\
 & & & & & 1.04e-12 & G017.8-02.6 &  & & & \\
 & & & & & 4.01e-12 & G018.9-01.1* &  & & & \\
 177 & (35.64, 0.01) & 0.36 & 36 & 1.86 & 9.51e-12 & G036.6-00.7 & J1852+0050u* & KM2A* & 0.85 & 1.85  \\
 & & & & & & & & WCDA & 0.64 & 1.07 \\
 & & & & & & &  {\bf J1857+0245} & WCDA & 0.24 & 0.44 \\
 & & & & & & &  {\bf J1857+0203u} & KM2A & 0.28 & 0.46 \\
 & & & & & & & & WCDA & 0.19 & 0.60 \\
 266 & (22.10, -1.06) & 0.16 & 114 & 4.21 & 4.65e-12 & G021.0-00.4 & J1831-1007u* & WCDA & 0.78 & 1.06  \\
 & & & & & 4.41e-12 & G021.6-00.8 &  & KM2A & 0.26 & 1.05 \\
 & & & & & 2.01e-15 & \textbf{G023.3-00.3} & & & & \\
230 & (21.97, -0.29) & 0.32 & 52 & 3.57 & 9.05e-12 & G021.6-00.8 & J1831-1007u* & WCDA & 0.78 & 0.40 \\
 & & & & & & & & KM2A & 0.26 & 0.26  \\
 & & & & & & & J1831-1028 & KM2A & 0.94 & 0.64 \\
301 & (21.78, -0.40) & 0.11 & 216 & 5.11 & 5.81e-12 & G021.0-00.4 & J1831-1007u* & WCDA & 0.78 & 0.33  \\
 & & & & & 2.18e-12 & G021.6-00.8 & & KM2A & 0.26 & 0.38 \\
 & & & & & & & J1831-1028 & KM2A & 0.94 & 0.46 \\
237 & (16.24, -1.02) & 0.17 & 139 & 4.33 & 6.08e-12 & G016.4-00.5 & J1825-1418 & WCDA & 0.81 & 1.01  \\
 & & & & & & & & KM2A & 0.81 & 1.31 \\
289 & (21.25, 0.02) & 0.09 & 244 & 4.72 & 4.05e-12 & G021.0-00.4 & J1831-1007u* & WCDA & 0.78 & 0.38 \\
 & & & & & 1.77e-12 & G021.6-00.8 & & KM2A & 0.26 & 0.64 \\
 & & & & & 2.24e-13 & G021.8-00.6 & J1831-1028 & KM2A & 0.94 & 0.39 \\
 173 & (18.32, 2.51) & 0.45 & 27 & 2.32 & 5.99e-12 & G019.1+00.2 & {\bf J1813-1245} & KM2A & 0.31 & 1.06 \\
 & & & & & & &  & WCDA & 0.32 & 1.04  \\
219 & (32.73, -0.42) & 0.14 & 134 & 2.91 & 3.81e-12 & G031.5-00.6 & {\bf J1850-0004u*} & WCDA & 0.46 & 0.50 \\
 & & & & & & & & KM2A & 0.21 & 0.39 \\
368 & (13.72, -0.14) & 0.17 & 21 & 8.10 & 9.32e-13 & G014.1-00.1 & {\bf J1814-1719u*} & WCDA & 0.71 & 0.34 \\
 & & & & & 8.58e-13 & G014.3+00.1 &  & KM2A & 0.27 & 1.03  \\
 & & & & & & & {\bf J1814-1636u} & KM2A & 0.68 & 0.53 \\
240 & (36.10, -0.14) & 0.26 & 90 & 3.40 & 1.64e-12 & G036.6-00.7 & {\bf J1857+0245} & WCDA & 0.24 & 0.06 \\
 & & & & & & & {\bf J1857+0203u} & KM2A & 0.28 & 0.70 \\
 & & & & & & &  & WCDA & 0.19 & 0.78 \\
 & & & & & & & {\bf J1858+0330 } & KM2A* & 0.43 & 0.77  \\
 & & & & & & &  & WCDA & 0.52 & 1.01 \\
    \end{tabular}
    \caption{List of molecular clouds illuminated from escaping CRs in SNRs, assuming a type II SN scenario, reported in descending order of predicted gamma-ray flux within this model. Clouds are selected from the \cite{2016ApJ...822...52Rice} catalog, while SNRs from the SNRcat compiled in \citet{snrcat}. Table columns report (from left to right) the Galactic coordinates of the cloud $(l,b)$, angular radius $R_{\rm cloud}$, number density $n$, distance from Earth $D$, predicted gamma-ray integral flux above 10\,TeV due to each illuminating SNR individually, and name of the illuminating SNRs as listed in order of descending contributions to the total flux of each cloud. On the right side of the vertical line, we report the spatially coincident sources from the 1st LHAASO catalog and their properties, including the name of the instrument reporting its measurement (LHAASO WCDA and/or KM2A), angular radius of the UHE emission (39\% containment) and angular separation from the cloud - note that, in case of more than one coincident LHAASO source, the ordering is independent of the order of coincident SNRs per cloud. SNRs reported in bold have a confirmed distance estimation, whilst 1LHAASO sources in bold indicate that their nature is as yet unidentified. }
    \label{tab:coincident}
    
\end{table*}


\begin{table*}
    \small
    \centering
    \begin{tabular}{c|cccccc|llcc}
    Cloud & ($l$,$b$) & R$_{cloud}$ & $n$ & D & $\phi_\gamma>10\,{\rm TeV}$ & SNR & 1LHAASO & inst. & R$_{39}$ & Sep. \\
     & ($^\circ$,$^\circ$) & ($^\circ$) & (cm$^{-3}$) & (kpc) & (TeV\,cm$^{-2}$s$^{-1}$) & & & & ($^\circ$) & ($^\circ$) \\
     \hline
201 & (14.22, -0.20) & 0.17 & 332 & 3.59 & 8.88e-11 & G012.7-00.0 & {\bf J1814-1719u*} & WCDA & 0.71 & 0.85  \\
 & & & & & 6.63e-11 & G014.3+00.1 &  & KM2A & 0.27 & 1.54 \\
 & & & & & & & {\bf J1814-1636u} & KM2A & 0.68 & 0.53 \\
 151 & (16.97, 0.53) & 0.46 & 90 & 2.08 & 1.16e-10 & G016.4-00.5 & J1825-1418 & WCDA & 0.81 & 1.37 \\
 & & & & & 1.65e-11 & G019.1+00.2 & J1825-1418 & KM2A & 0.81 & 1.27 \\
214 & (18.40, -0.27) & 0.24 & 163 & 3.60 & 5.37e-11 & G017.0-00.0 & J1825-1418 & WCDA & 0.81 & 1.29 \\
 & & & & & 7.06e-11 & G017.4-00.1 &  & KM2A & 0.81 & 0.97 \\
 & & & & & 8.13e-13 & G019.1+00.2  & {\bf J1825-1256u} & KM2A & 0.20 & 0.12 \\
 & & & & & & &  & WCDA & 0.24 & 0.18 \\
 & & & & & & & J1825-1337u & KM2A & 0.18 & 0.60 \\
 & & & & & &  &  & WCDA & 0.17 & 0.72 \\
 155 & (14.14, -0.59) & 0.18 & 527 & 2.09 & 1.23e-10 & G016.4-00.5 & {\bf J1814-1719u*} & WCDA & 0.71 & 0.94 \\
 & & & & & & & & KM2A & 0.27 & 1.57 \\
 & & & & & & & {\bf J1814-1636u} & KM2A & 0.68 & 0.88 \\
208 & (16.61, -0.38) & 0.31 & 141 & 3.65 & 8.95e-11 & G017.4-00.1 & J1825-1418 & WCDA & 0.81 & 0.75 \\
 & & & & & & &  & KM2A & 0.81 & 0.91 \\
225 & (13.94, 0.22) & 0.07 & 479 & 4.10 & 2.08e-11 & G012.7-00.0 & {\bf J1814-1719u*} & WCDA & 0.71 & 0.60 \\
 & & & & & 2.08e-11 & G014.1-00.1 &  & KM2A & 0.27 & 1.24  \\
 & & & & & 2.43e-11 & G014.3+00.1 & {\bf J1814-1636u} & KM2A & 0.68 & 0.12  \\
177 & (35.64, 0.01) & 0.36 & 36 & 1.86 & 6.85e-12 & G036.6-00.7 & J1852+0050u* & KM2A* & 0.85 & 1.85 \\
 & & & & & 2.38e-11 & G036.6+02.6 &  & WCDA & 0.64 & 1.07 \\
 & & & & & & & {\bf J1857+0245} & WCDA & 0.24 & 0.44  \\
 & & & & & & & {\bf J1857+0203u} & KM2A & 0.28 & 0.46 \\
 & & & & & & & & WCDA & 0.19 & 0.60 \\
 230 & (21.97, -0.29) & 0.32 & 52 & 3.57 & 2.08e-11 & G021.0-00.4 & J1831-1007u* & WCDA & 0.78 & 0.40 \\
 & & & & & 7.92e-12 & G021.6-00.8 & & KM2A & 0.26 & 0.26 \\
 & & & & & & & J1831-1028 & KM2A & 0.94 & 0.64 \\
260 & (14.61, -0.34) & 0.13 & 90 & 4.68 & 1.03e-11 & G014.1-00.1 & {\bf J1814-1636u} & KM2A & 0.68 & 0.85 \\
 & & & & & 1.31e-11 & G014.3+00.1 & & & & \\
 190 & (34.99, -0.96) & 0.38 & 126 & 2.62 & 9.98e-12 & G036.6-00.7 & J1852+0050u* & KM2A* & 0.85 & 1.65 \\
 & & & & & & &  & WCDA & 0.64 & 0.93 \\
 & & & & & & & {\bf J1857+0203u} & KM2A & 0.28 & 0.72 \\
 & & & & & & &  & WCDA & 0.19 & 0.61 \\
    \end{tabular}
    \caption{As in table \ref{tab:coincident} for the type Ia SNe scenario. }
    \label{tab:coincidentIA}
\end{table*}

\section{Global comparison of Galactic population}
\label{sec:population}

In \cite{2021MNRAS.503.3522Mitchell}, gamma-ray flux predictions were derived for molecular clouds identified by \cite{2016ApJ...822...52Rice}\footnote{Note that in this work cloud IDs refer to the ``cloud index'' in the FITS file from \cite{2016ApJ...822...52Rice}, whereas in \cite{2021MNRAS.503.3522Mitchell} we referred to the clouds in the order of their listing in the machine-readable table 3 of \cite{2016ApJ...822...52Rice}.} from CO data of \cite{Dame01}, under the scenario of CR acceleration by the SNRs listed in the SNR catalogue (SNRcat), as compiled by \cite{snrcat}. Using an up-to-date version of SNRcat,\footnote{\url{ http://snrcat.physics.umanitoba.ca}, accessed 26/08/2024} we paired SNRs to clouds based on their coordinates and distance. \cite{2016ApJ...822...52Rice} provided either an unambiguous distance or a preferred distance estimate (near or far) for each cloud, which we used for this study. Where a distance estimate for an SNR is available, this is required to be consistent with the distance of the cloud (within errors); otherwise each SNR is assumed to be located at the same distance from Earth as the respective cloud, for the purposes of our model. Each pair must have a physical separation no larger than 100\,pc to be taken into account. 

Although there are considerable uncertainties inherent in the modelling, related (for example) to the transport properties of the CRs nearby the accelerators or to the SNR progenitor, we previously adopted only a type II SNR scenario for particle acceleration. This assumption was mostly driven by the fact that dense clouds are located nearby core collapse SNe, because these systems explode near or in the same clouds where they formed, differently from the thermonuclear case. We follow up on those studies by exploring here spatial and spectral coincidences with LHAASO sources. Moreover,
to account for the possibility that the target cloud is unrelated to the environment of the coupled SNR, we also explore here the scenario of type Ia explosions illuminating clouds. 

Limited information is provided in SNR catalogues about the progenitor origin,  such that for most of the considered SNRs we explore here both scenarios, either type II or type Ia SNe respectively.  For the cases when a PWN is explicitly reported (SNRcat) in association to the SNR (i.e. a plerionic type), we instead consider only the type II origin. 
In particular, figure \ref{fig:galplane} compares the results of our model obtained in the scenario of type II explosions, with the locations of LHAASO sources from the first catalog \citep{2024ApJS..271...25Cao_1lhaaso} indicated by red circles, and with the contours from the H.E.S.S. Galactic Plane Survey (HGPS) \citep{2018A&A...612A...1HGPS}. By contrast, figure \ref{fig:galplaneIa} refers to the scenario where the particle illuminating flux is provided by type Ia SNe.
Only a small section of the Galactic Plane is covered by both instruments, with LHAASO data extending towards the Northern sky whilst HGPS data coverage continues towards the Galactic centre and beyond into the Southern sky, highlighting the need for a complementary extensive air shower array in the Southern Hemisphere such as the planned SWGO \citep{swgo}.  \\

The situation as shown in Figure \ref{fig:galplane} is relatively complex. Along spiral arms and towards lower Galactic longitudes, there is an increasing density of molecular clouds and gamma-ray sources overlapping along the line-of-sight, that further complicates the identification of counterpart objects. To identify regions that require more detailed studies, we searched for illuminated clouds that are coincident with 1LHAASO sources along the line-of-sight. Of particular interest are systems that do not have a clear physical counterpart (UNID), as in several cases the total emission may be contributed to by an illuminated cloud scenario.

We consider a cloud to be coincident with a 1LHAASO source if the angular separation between their best-fit positions is less than the sum of their radii (implicitly assuming spherical symmetry). The 39\% containment radius is used for the 1LHAASO sources, which is conservative with respect to the total extent of the emission. 
Table \ref{tab:coincident} summarises the coincident clouds and LHAASO sources that were identified, where a coincidence is evaluated assuming the most positive scenario where the cloud and the LHAASO source are at the same distance from Earth, given that the distance to LHAASO sources cannot be determined solely from the gamma-ray emission. The same table also provides the putative physical association for the gamma-ray source, as indicated in \citep{2024ApJS..271...25Cao_1lhaaso}. From this information, it emerges that the majority of overlapping sources are either energetic pulsar environments or unidentified in nature. Most importantly, the table further provides an identifier for the illuminating SNRs found in our model in accordance with that from \cite{2016ApJ...822...52Rice}: as previously mentioned, in many cases each cloud is actually illuminated by more than one SNR. Where the distance to an SNR is known, this SNR is indicated in bold, indicating that the spatial match between SNR and cloud is confident. 
Table \ref{tab:coincidentIA} lists clouds illuminated by SNRs under the type Ia scenario. None of the SNRs are listed as containing pulsars in \citet{snrcat}, that would otherwise exclude a type Ia origin. 
Although gamma rays from illuminated clouds may contribute to the total flux of several LHAASO sources, the most promising cases are those of unidentified sources for which no counterpart or physical explanation is yet known, on which we will focus in the next section.

\section{Detailed comparison of specific sources}
\label{sec:sources}

This section is dedicated to the investigation of spectral properties of unidentified sources in the 1LHAASO source catalog and the clouds with which they are spatially coincident, as reported in table \ref{tab:coincident}.
For the purposes of this section, we stress that the following results do not show any fit to data, but rather a simple spectral comparison among 1LHAASO sources and the modelling results for coincident interstellar clouds illuminated by SNR-escaping CRs with the aforementioned set of assumptions.  In fact, a linear scaling between gamma-ray flux and CR conversion efficiency applies, such that fitting the overall normalisation of the expected gamma-ray flux would provide this parameter individually for each source. It is interesting to note that the standard 10\% assumption already ensures a reasonable flux level compared to data. 
As we are considering two alternative scenarios, due to type II and type Ia SNe respectively as responsible for the CR flux, we treat the type II origin as the baseline case. The typical effect of the type Ia origin is to increase the total flux reaching the cloud from the same SNRs. In a few cases, a type Ia scenario leads more SNRs to contribute, i.e. CRs only reach certain clouds from SNRs under the type Ia scenario. This also leads to a few cases where the cloud is only significantly illuminated for a type Ia scenario. This is a consequence of the decreased $t_{\rm sed}$ for type Ia SNe, with respect to the type II scenario in the simplifying assumption of uniform circumstellar density profile. 

\begin{figure*}
    \centering
    \begin{overpic}[width=0.9\columnwidth]{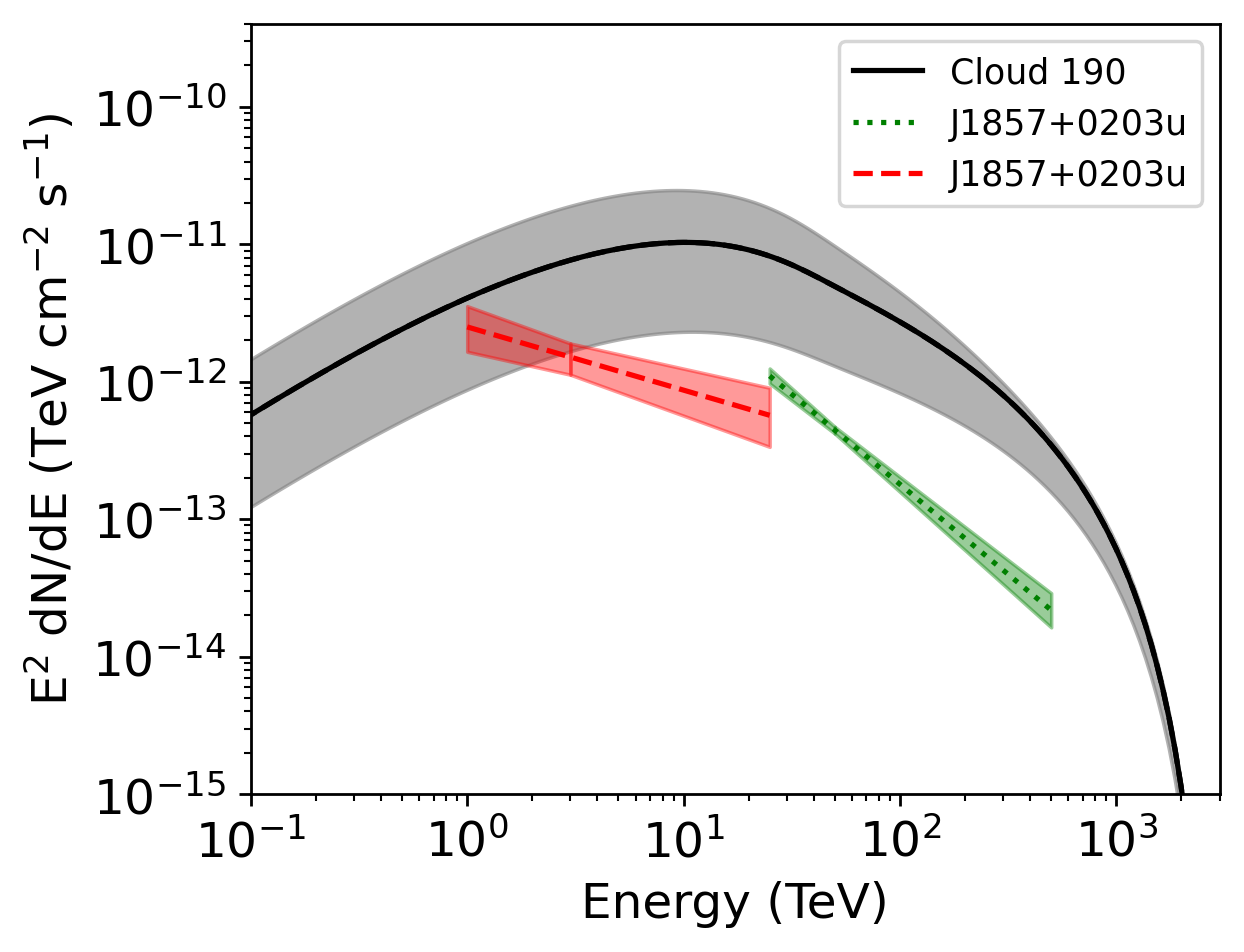}  
    \put(25,68){\small type II}
    \end{overpic}
    \begin{overpic}[width=0.9\columnwidth]{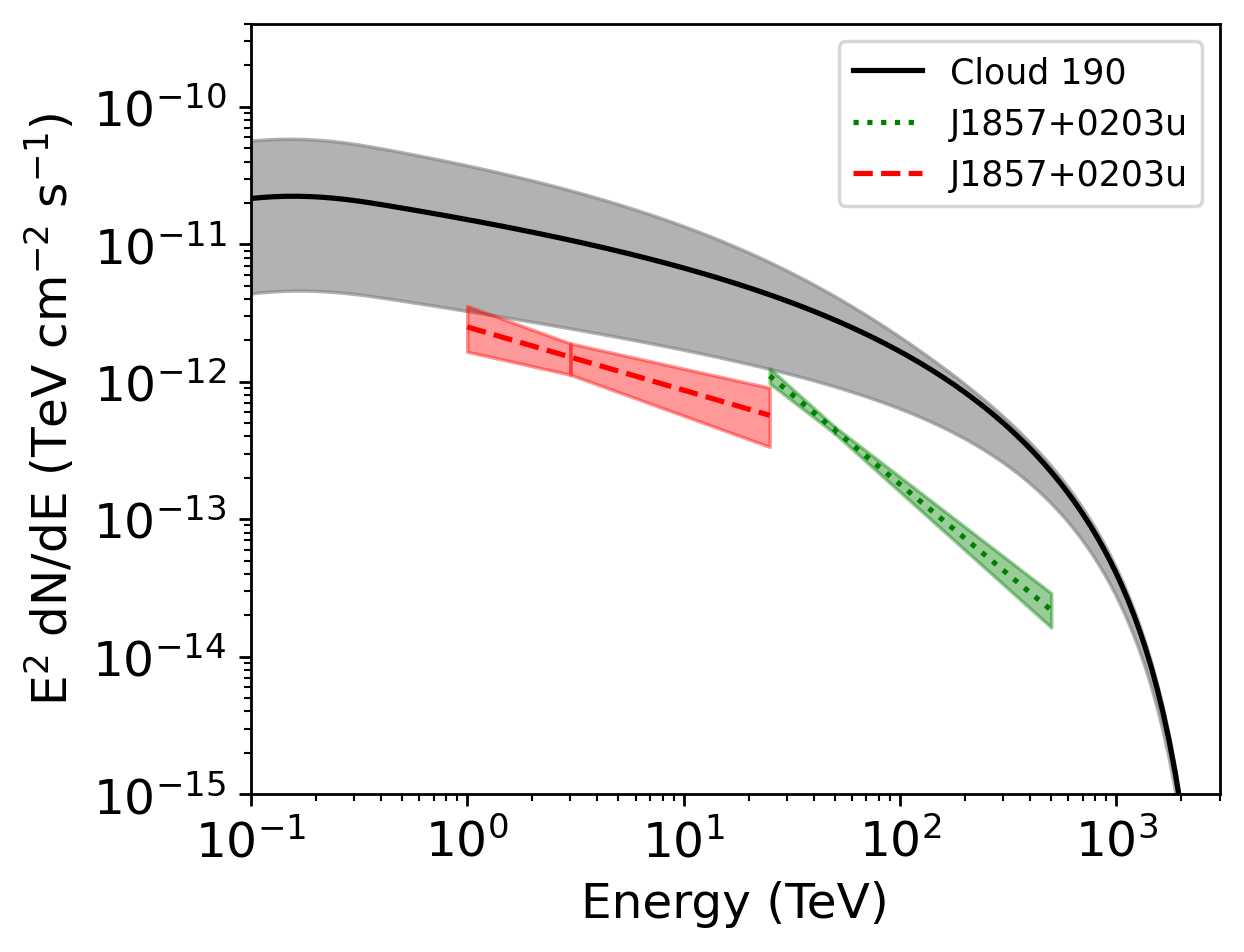}
    \put(25,68){\small type Ia}
    \end{overpic}
    \begin{overpic}[width=0.9\columnwidth]{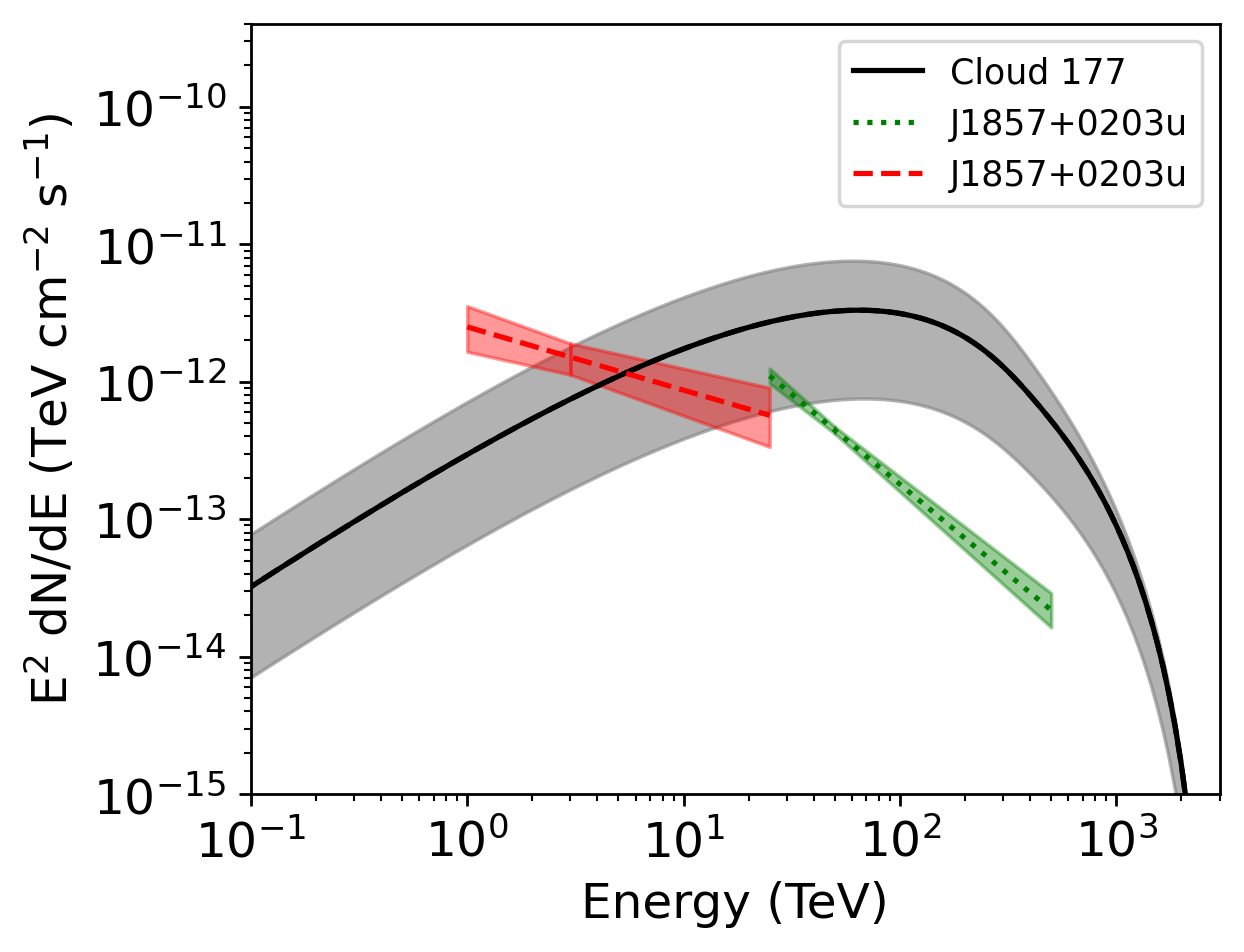}
    \put(25,68){\small type II}
    \end{overpic}
    \begin{overpic}[width=0.9\columnwidth]{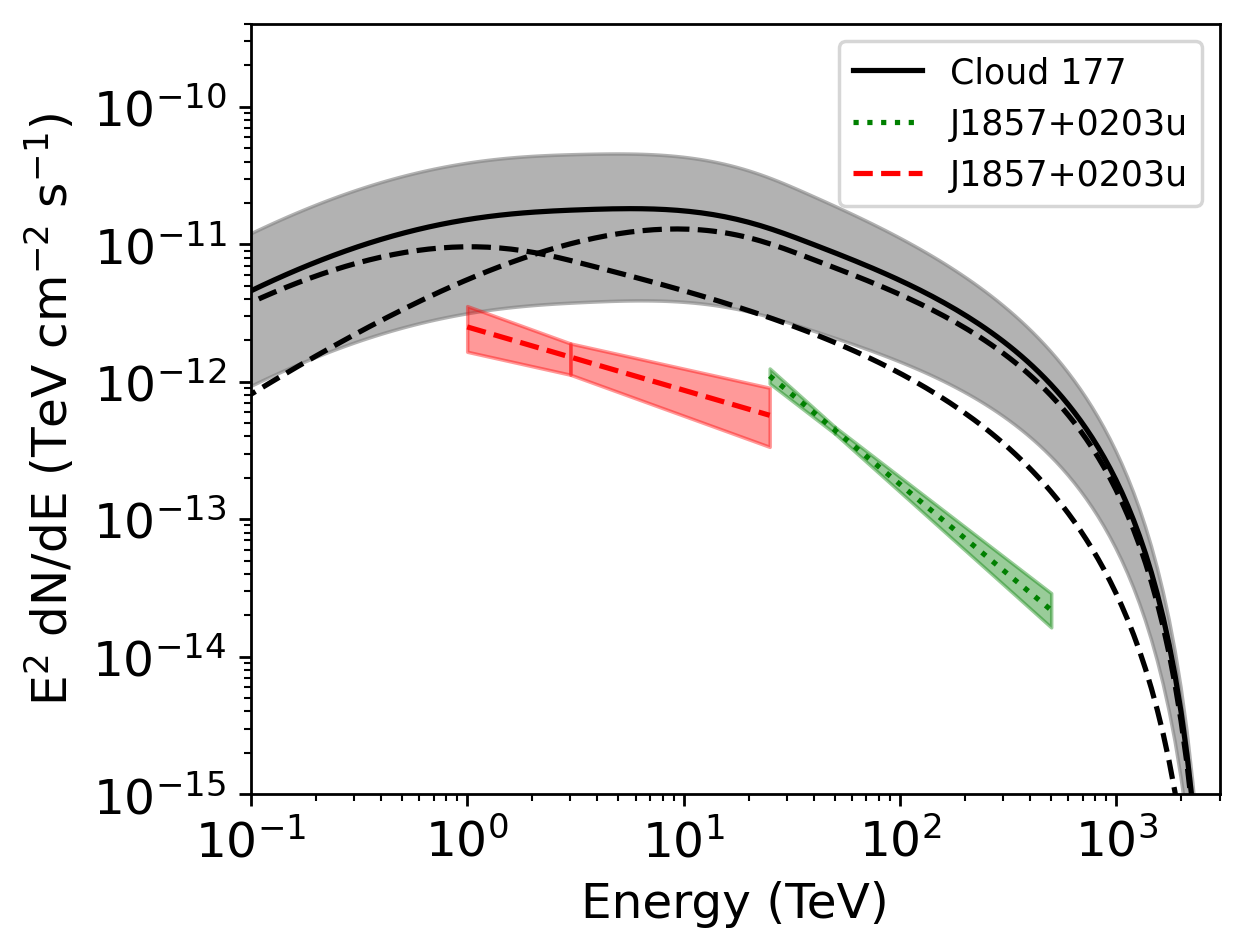}
    \put(25,68){\small type Ia}
    \end{overpic}
    \begin{overpic}[width=0.9\columnwidth]{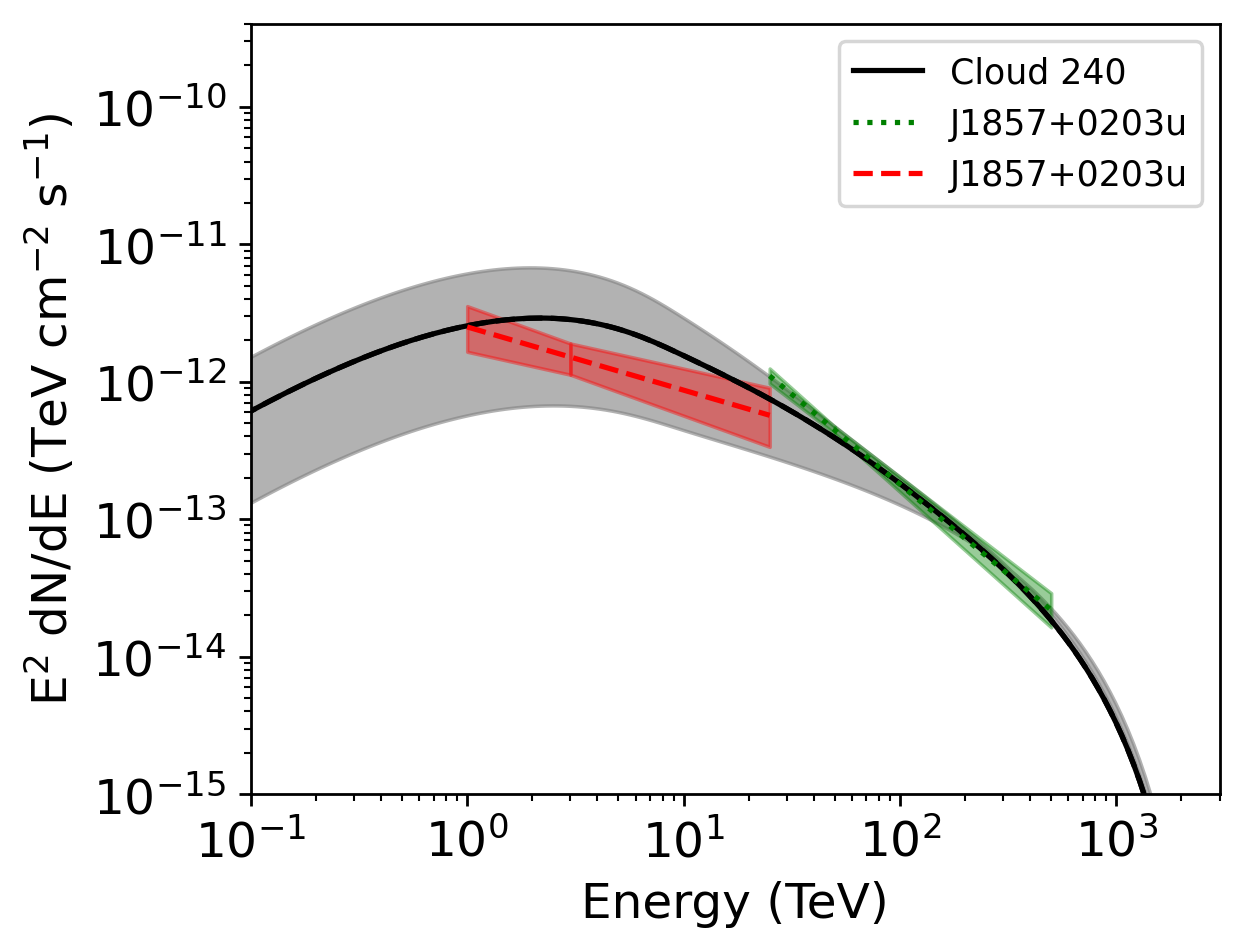}
    \put(25,68){\small type II}
    \end{overpic}

    \caption{1LHAASO\,J1857+0203u compared to clouds\,190, 177 and 240 at $(l,b)=(34.99^\circ,-0.96^\circ)$,  $(l,b)=(35.64^\circ,0.01^\circ)$ and $(l,b)=(36.10^\circ,-0.14^\circ)$ . Where more than one SNR contributes, the individual contributions are indicated by dashed lines, with the solid line indicating the total flux from the cloud. Power law spectra shown with a red dashed or green dotted line correspond to measurements by LHAASO WCDA or KM2A respectively. The SN scenario is also labelled on the plot.}
    \label{fig:c190_j1857}
\end{figure*}

\begin{figure*}
    \centering
    \begin{overpic}[width=0.9\columnwidth]{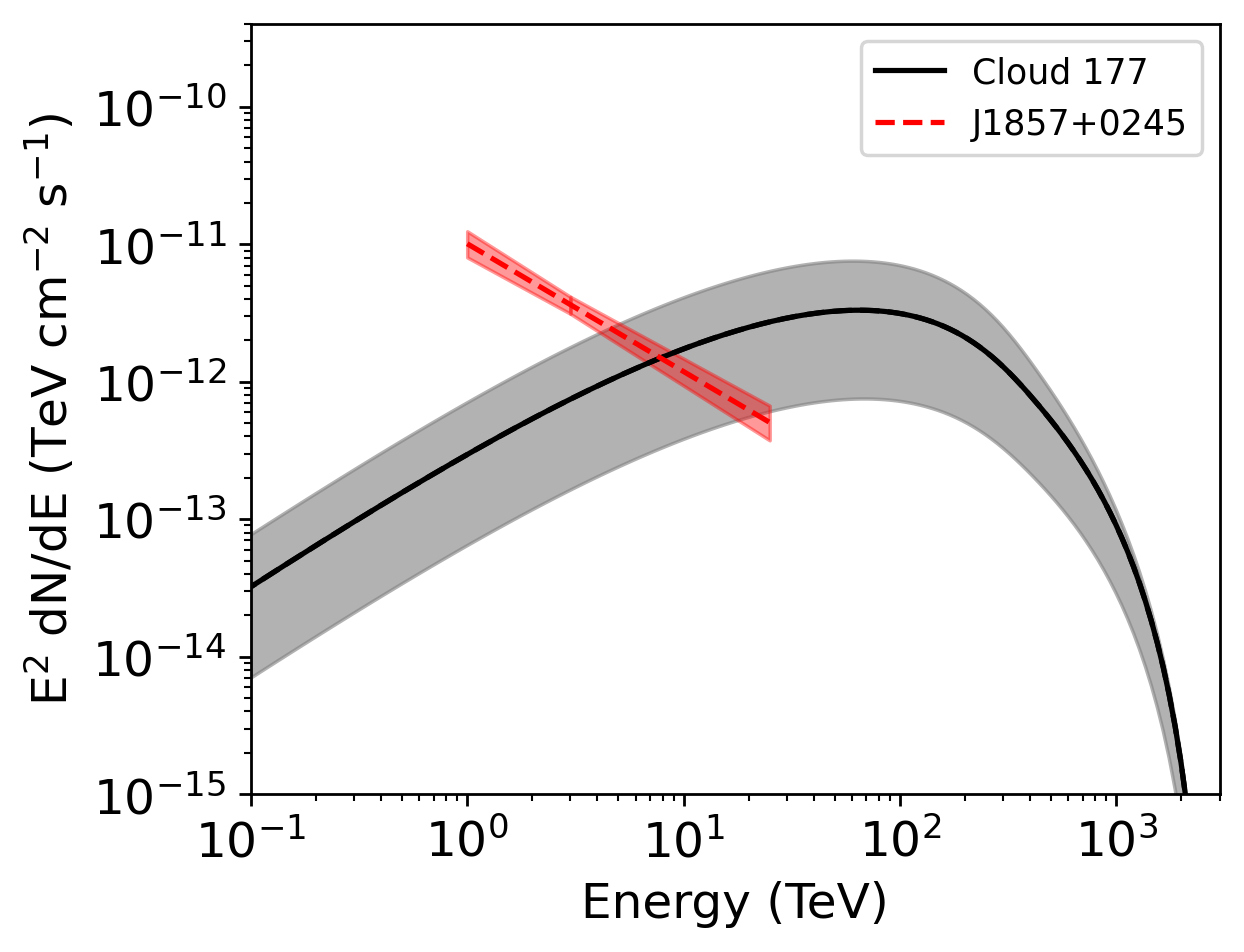}
    \put(25,68){\small type II}
    \end{overpic}
    \begin{overpic}[width=0.9\columnwidth]{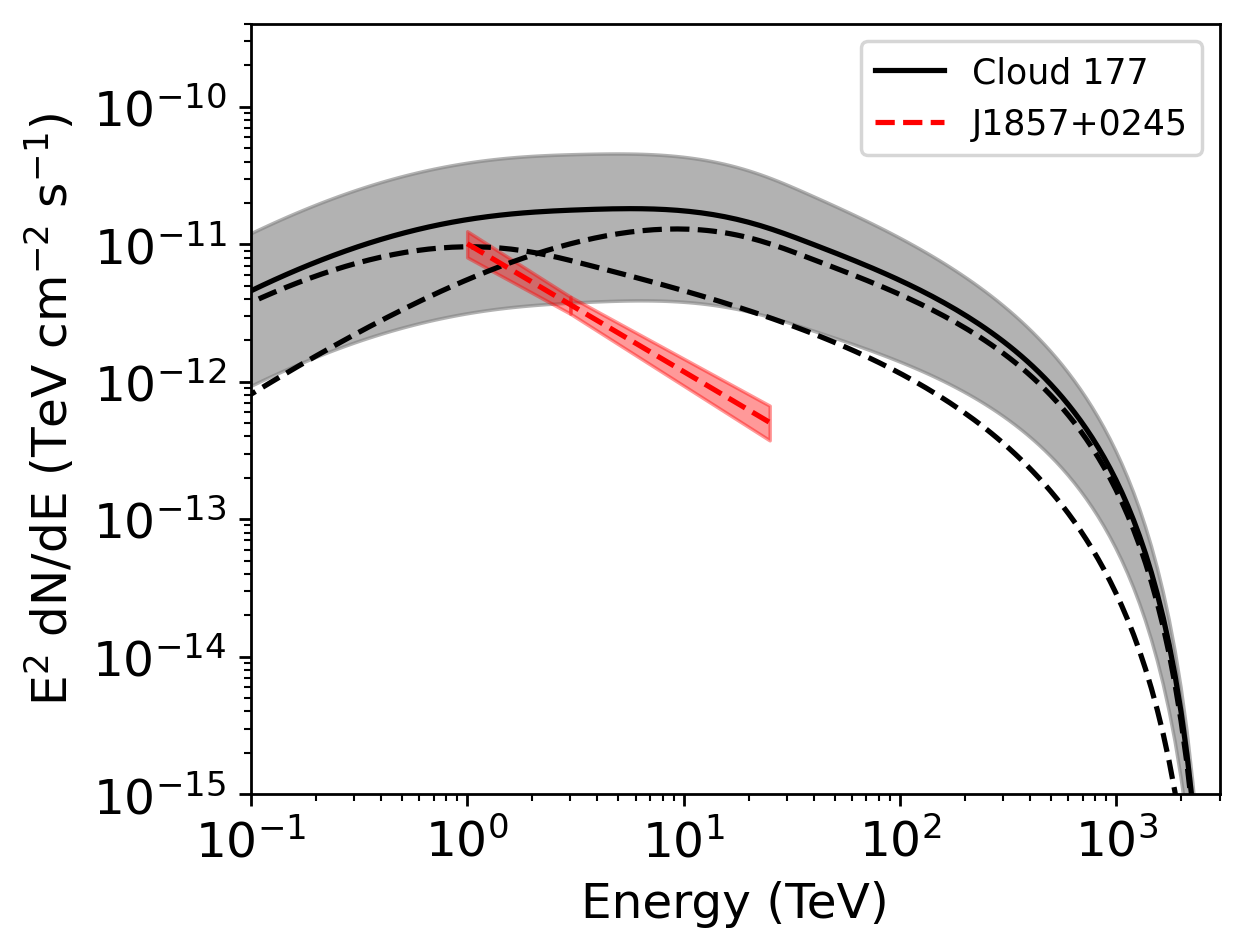}
    \put(25,68){\small type Ia}
    \end{overpic}
    \begin{overpic}[width=0.9\columnwidth]{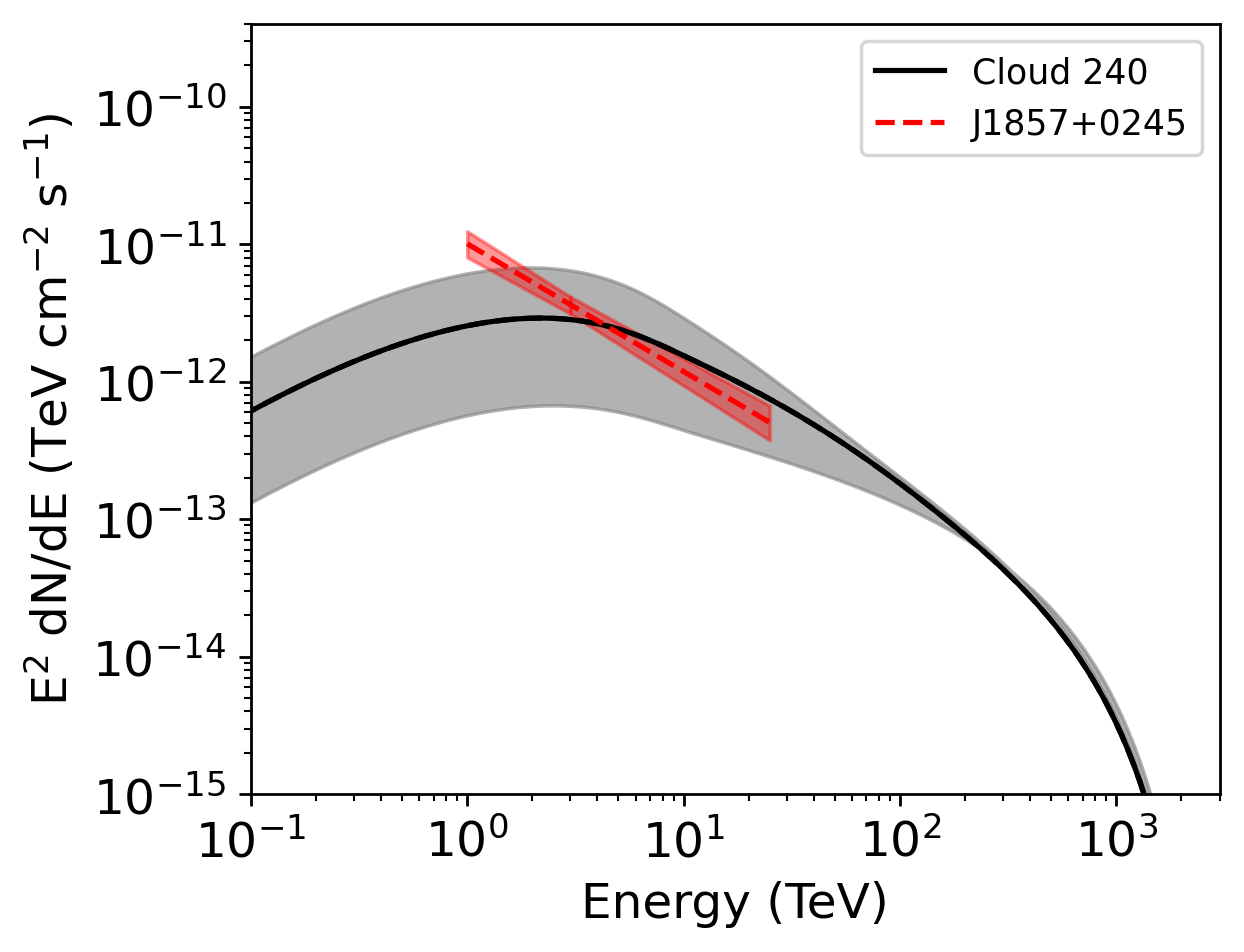}
    \put(25,68){\small type II}
    \end{overpic}
    \caption{As in figure \ref{fig:c190_j1857} for 1LHAASO\,J1857+0245 with clouds\,177 and 240, located at $(l,b)=(35.64^\circ,0.01^\circ)$ and $(l,b)=(36.10^\circ,-0.14^\circ)$ respectively. }
    \label{fig:c177_j1857}
\end{figure*}

In all cases shown from Figure \ref{fig:c190_j1857} onwards, shaded bands indicate a conservative estimation of the uncertainties inherent in our model. The breadth of the shaded band is given by variation in the normalisation of the diffusion coefficient $D_0$ by a factor $\sim10$, of the level of turbulent suppression $\chi$ by a factor $\sim10$, the presence or absence of a contribution from the diffuse Galactic CR flux, a minimum and maximum estimate of the SNR distance if available (otherwise we assumed a 20\,\% uncertainty on the distance) and the minimum and maximum estimates for the SNR age (if available).

We discuss in this section all coincidences of clouds with unidentified 1LHAASO sources, as listed in tables \ref{tab:coincident} and \ref{tab:coincidentIA}. In several cases multiple SNRs contribute to form the total gamma-ray flux - we only consider those that contribute at a level $\geq 1\%$, which are also listed in table \ref{tab:coincident}.

\subsection{1LHAASO J1857+0203u}

Figure \ref{fig:c190_j1857} compares our model prediction for cloud\,190 to measured spectra for 1LHAASO\,J1857+0203u is associated to HESS\,J1858+020, an unidentified TeV source. Cloud\,190 at $(l,b)=(34.99^\circ,-0.96^\circ$) is illuminated by G036.6-0.7, about which comparatively little is known. 

Clouds 177 and 240 are also coincident with the same LHAASO source and possibly illuminated by the SNR G036.6-0.7. Cloud 240 under the type II SN scenario provides the best visual match to the LHAASO spectra, whilst the flux is overestimated in the other cases, implying that (e.g.) the assumed CR conversion efficiency is too high. 

\subsection{1LHAASO J1857+0245}
Additionally, clouds 177 and 240 are coincident with 1LHAASO\,J1857+0245, associated to HESS\,J1857+026, a source with complex morphology that has been preferentially described by MAGIC as comprised of two sources \citep{2014A&A...571A..96MagicJ1857}. As such, although the presence of an energetic pulsar PSR\,J1856+0245 implies that a PWN scenario may be viable for this source, it is unlikely for it to account for the entirety of the observed emission, such that an illuminated cloud could contribute to the observed flux. As in figure \ref{fig:c190_j1857}, cloud 240 again provides a reasonable match to the LHAASO spectrum. 
A physical connection between the SNR, TeV source and molecular clouds in the region is explored in more detail by \cite{2022ApJ...931..128ZhangJ1858}.

\subsection{1LHAASO J1813-1245}

\begin{figure}
    \centering
    \begin{overpic}[width=0.9\columnwidth]{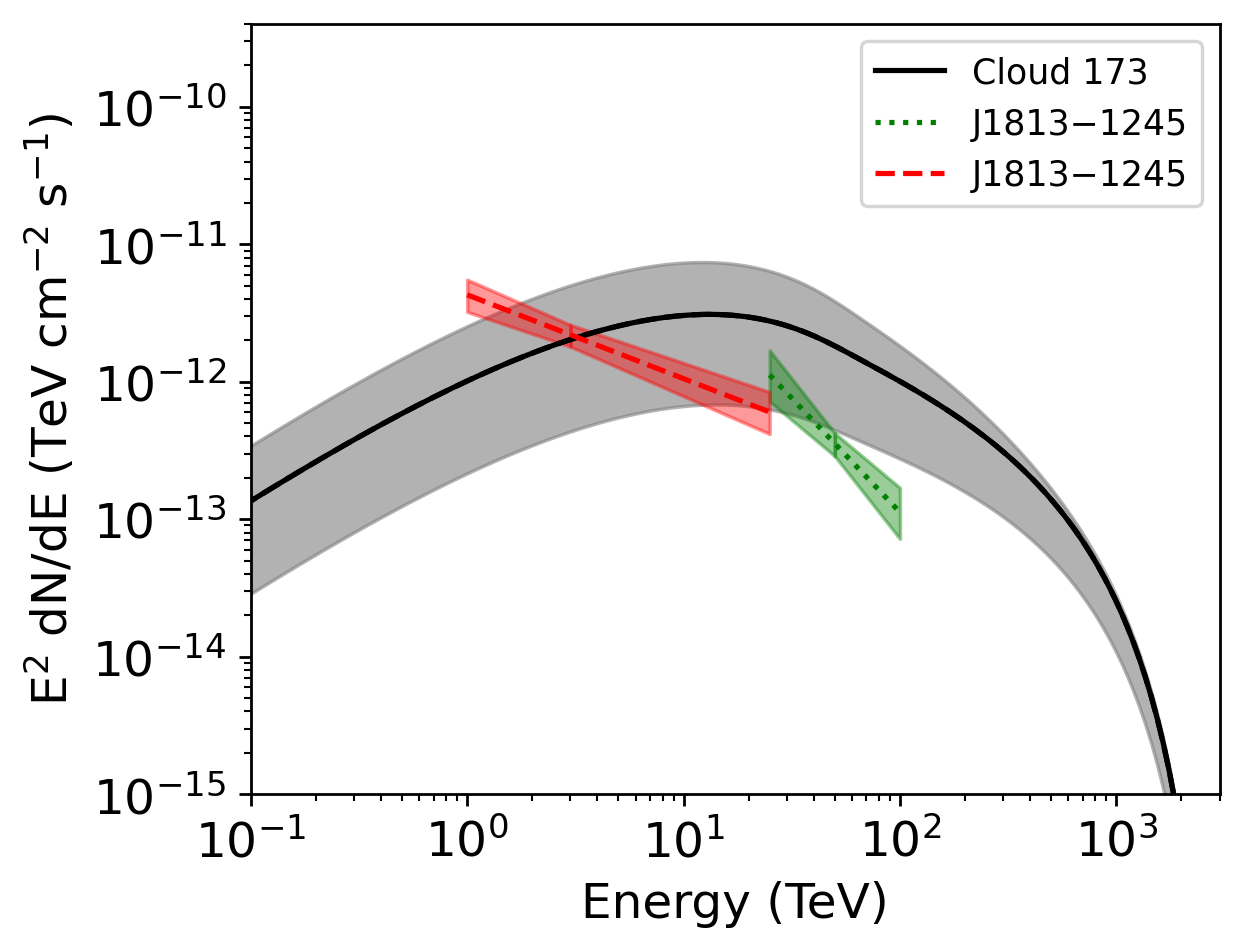}
    \put(25,68){\small type II}
    \end{overpic}
    \caption{As in figure \ref{fig:c190_j1857} for 1LHAASO\,J1813-1245 with cloud\,173 located at $(l,b)=(18.32^\circ,2.51^\circ)$. }
    \label{fig:c173_j1813}
\end{figure}

Figure \ref{fig:c173_j1813} shows 1LHAASO\,J1813-1245 coincident with HESS\,J1813-126, compared to cloud\,173. Although this source is formally unidentified, the PSR\,J1813-1246 provides a reasonable explanation for its emission, whilst there is a fairly large ($\gtrsim 2R_{\rm 39}$) spatial separation between cloud\,173 and 1LHAASO\,J1813-1245. As the illuminating SNR G019.1+0.2 is also poorly constrained, we consider the PWN scenario to be more likely in this case. 

\subsection{1LHAASO J1850-0004u*}
\begin{figure}
    \centering
    \begin{overpic}[width=0.9\columnwidth]{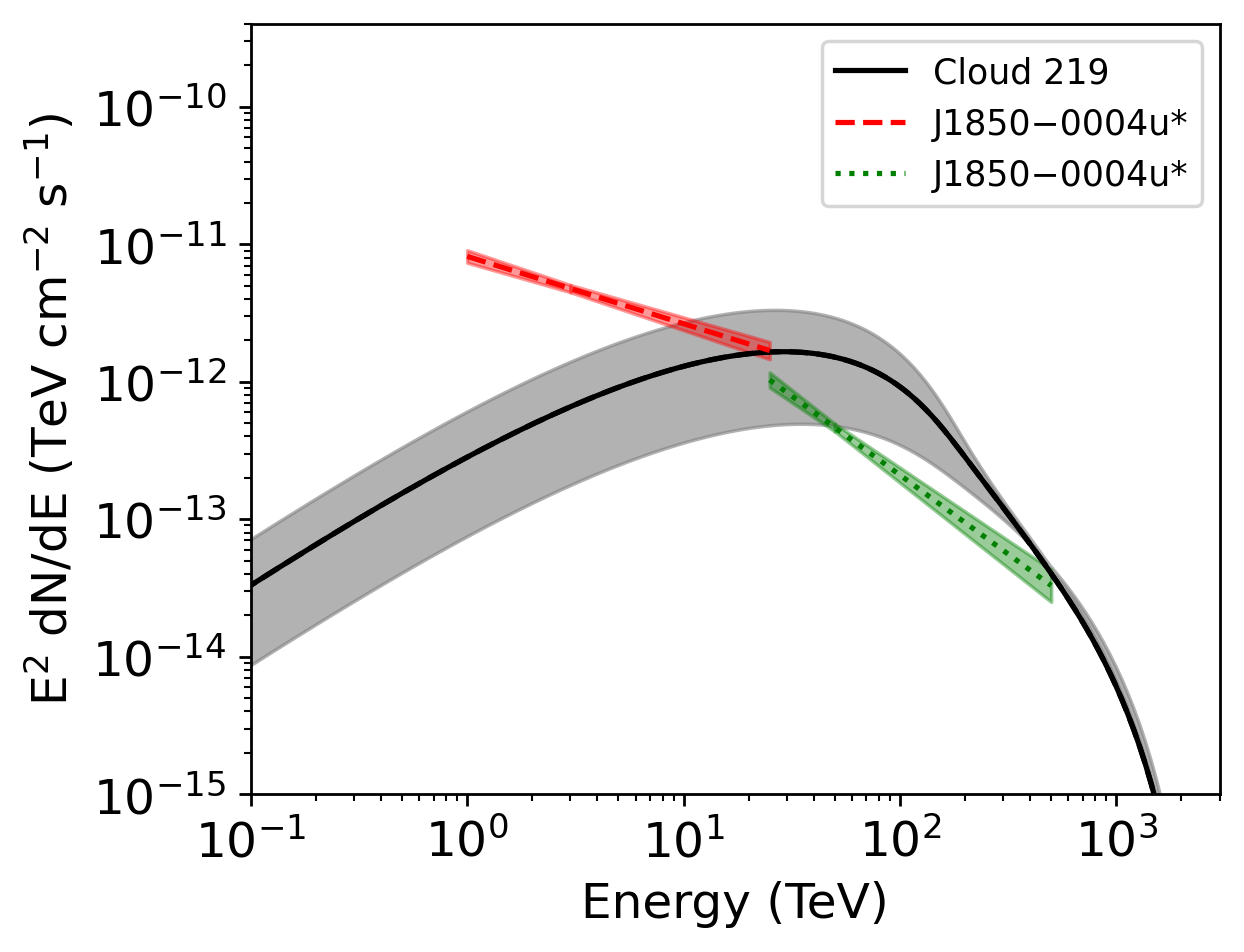}
    \put(25,68){\small type II}
    \end{overpic}
    \caption{As in figure \ref{fig:c190_j1857} for 1LHAASO\,J1850-0004u* with cloud 219 located at $(l,b)=(32.73^\circ,-0.42^\circ)$  under the type II scenario. }
    \label{fig:c219_j1850}
\end{figure}

Figure \ref{fig:c219_j1850} compares cloud\,219, putatively illuminated predominantly by the SNR G031.5-0.6, to 1LHAASO\,J1850-0004u*, a UHE gamma-ray source coincident with HESS\,J1852-000. This gamma-ray source remains unidentified, yet is tentatively associated with Kes\,78 \citep{2018A&A...612A...1HGPS}. G031.5-0.6 has counterpart gamma-ray emission detected by the Fermi-LAT, yet no confirmed TeV emission to date \citep{2016ApJS..224....8A_FermiSNR}. Although the properties of this SNR are not well-constrained, one can see that the predicted flux in the type II scenario is reasonably consistent with the measured flux for 1LHAASO\,J1850-0004u* at energies $\gtrsim10$\,TeV, providing a potential explanation for this source. As such, a more detailed investigation and modelling of this region is warranted. 

\subsection{1LHAASO J1814-1719u*}
\begin{figure*}
    \centering
    \begin{overpic}[width=0.9\columnwidth]{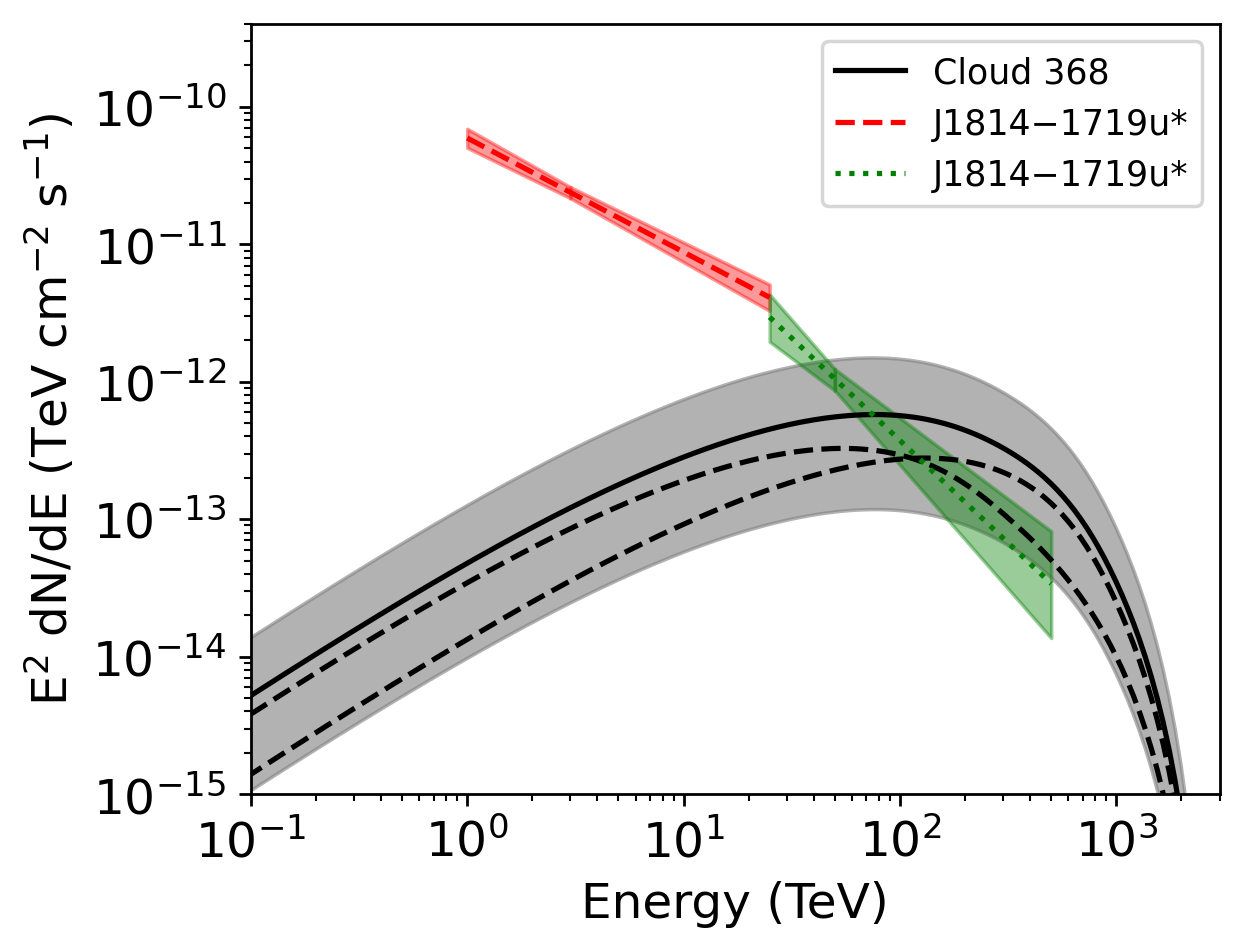}
    \put(25,68){\small type II}
    \end{overpic}
    \begin{overpic}[width=0.9\columnwidth]{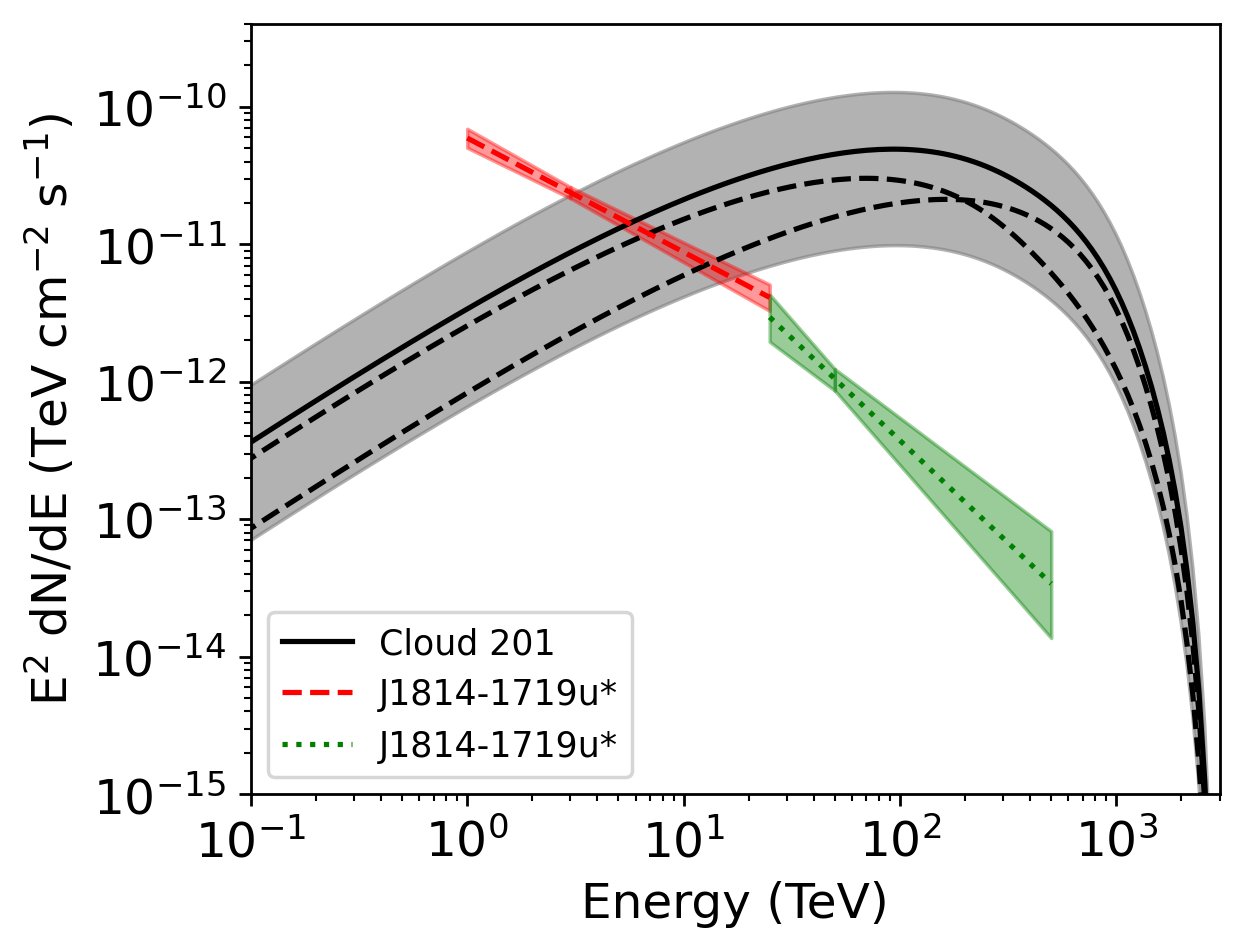}
    \put(25,68){\small type Ia}
    \end{overpic}
    \begin{overpic}[width=0.9\columnwidth]{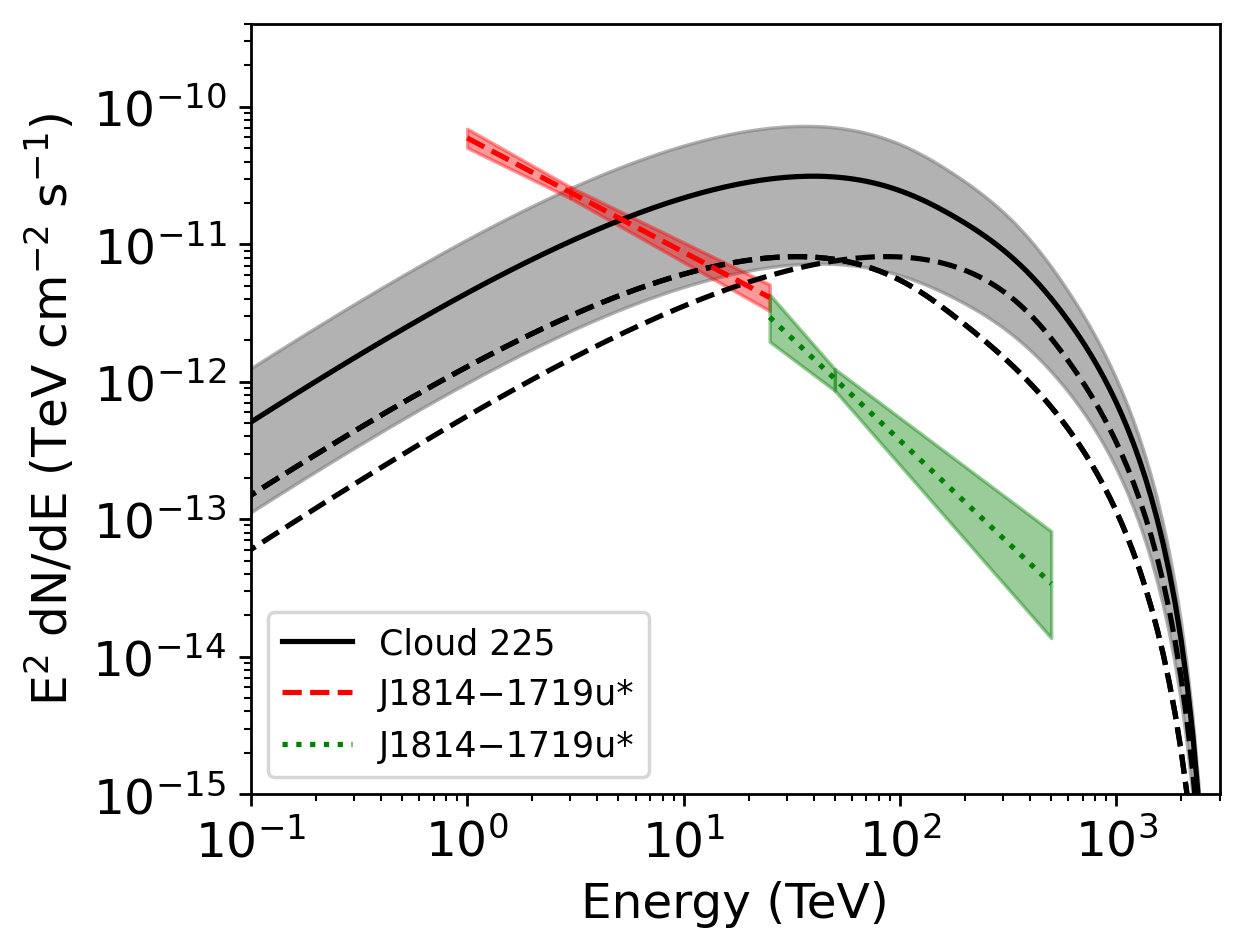}
    \put(25,68){\small type Ia}
    \end{overpic}
    \begin{overpic}[width=0.9\columnwidth]{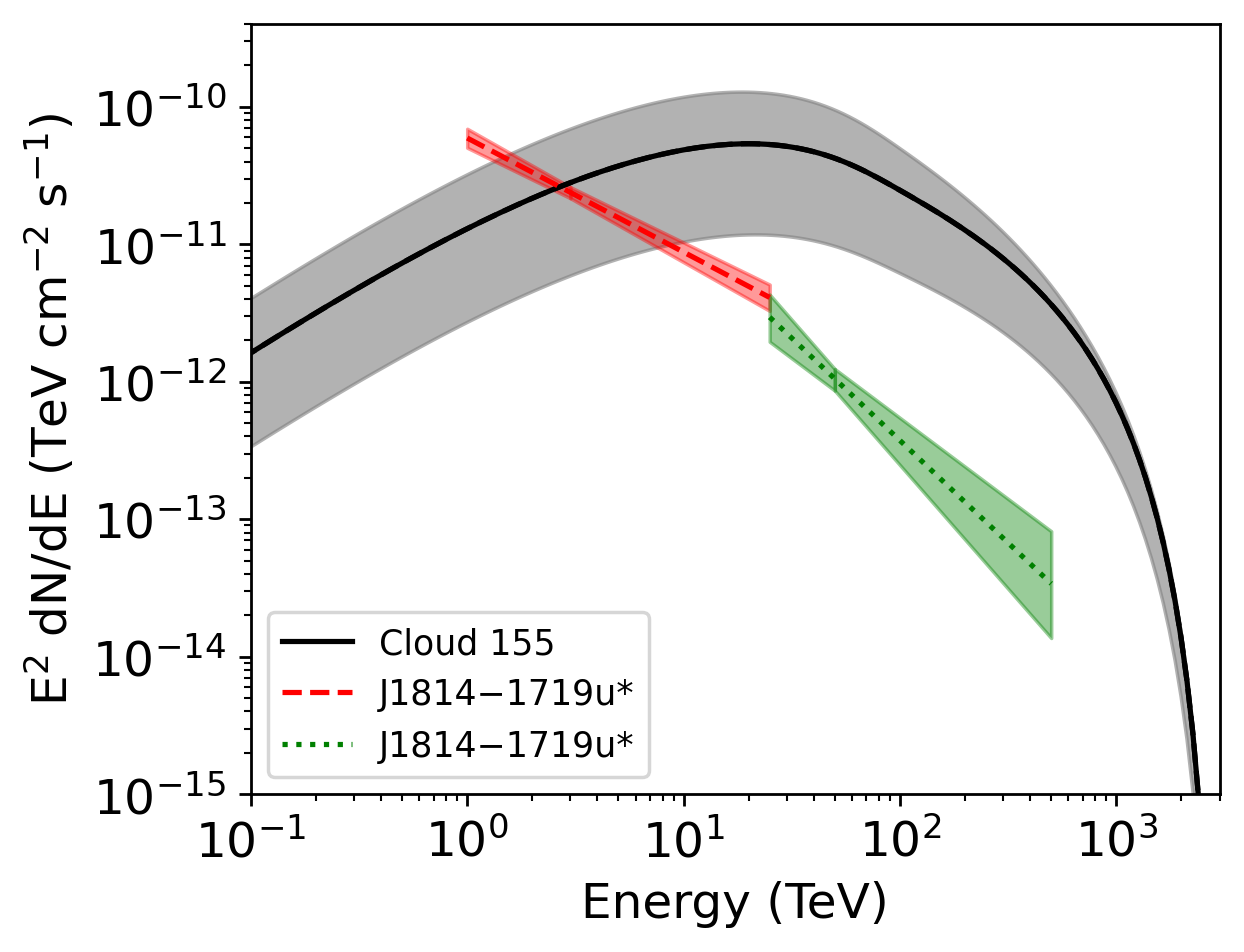}
    \put(25,68){\small type Ia}
    \end{overpic}
    \caption{As in figure \ref{fig:c190_j1857} for 1LHAASO\,J1814-1719u* with clouds 368, 201, 225 and 155 located at $(l,b)=(14.14^\circ,-0.59^\circ)$, $(l,b)=(14.22^\circ,-0.20^\circ)$, $(l,b)=(13.94^\circ,0.22^\circ)$ and $(l,b)=(14.14^\circ,-0.59^\circ)$ 
     respectively. }

    \label{fig:c368_j1814}    
\end{figure*}

Figure \ref{fig:c368_j1814} compares source 1LHAASO\,J1814-1719u*, of unidentified origin, to clouds\,368, 201, 225, and 155.  Cloud 368 is expected to be illuminated under the type II scenario by the SNR G014.1-0.1 and G014.3+0.1. Although the properties of this SNR are not well-constrained, the location of the cloud is well-contained within the emission region and  consistent with or below the measured flux.

\subsection{1LHAASO J1814-1636u}

\begin{figure*}
    \centering
    \begin{overpic}[width=0.9\columnwidth]{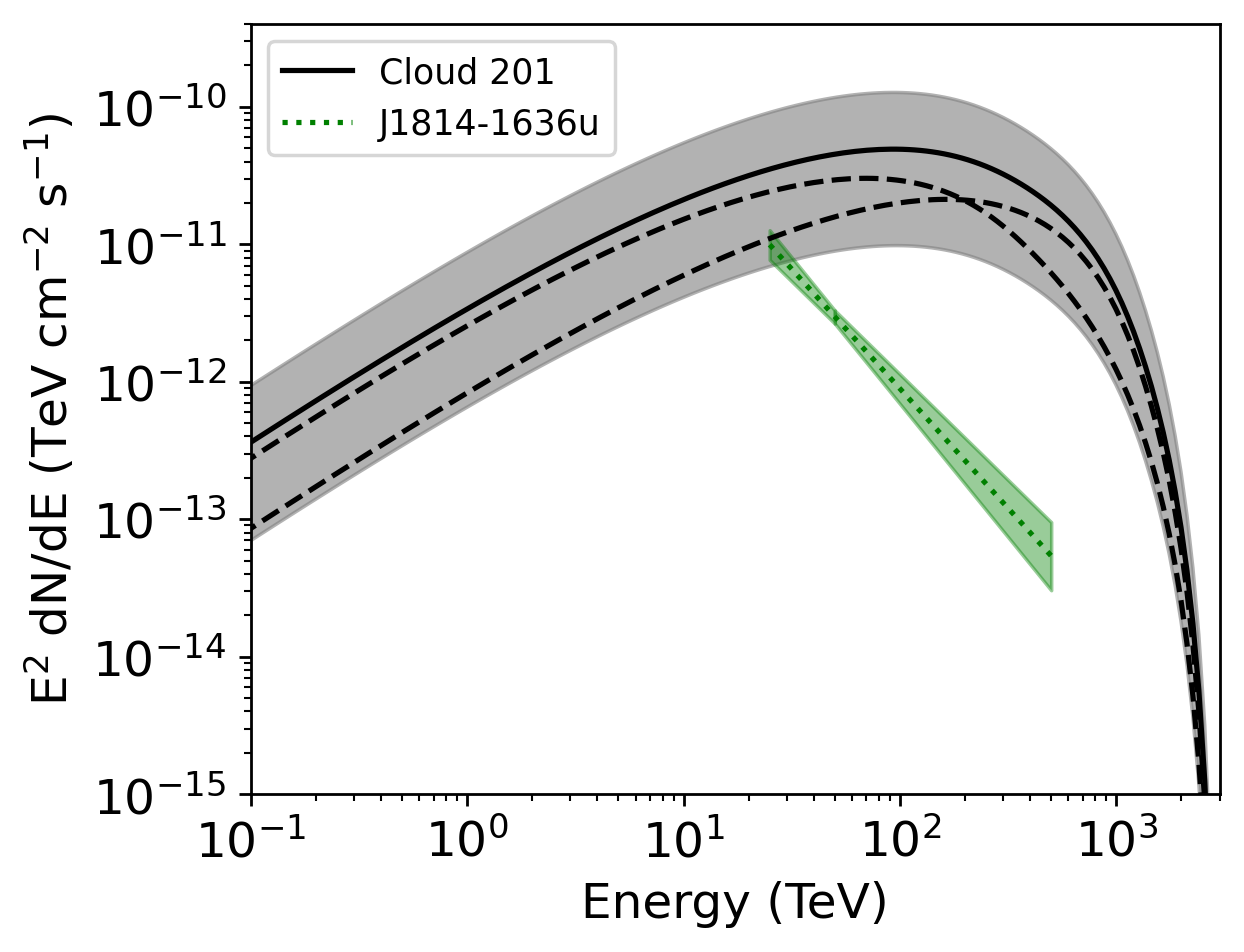}
    \put(80,68){\small type Ia}
    \end{overpic}
    \begin{overpic}[width=0.9\columnwidth]{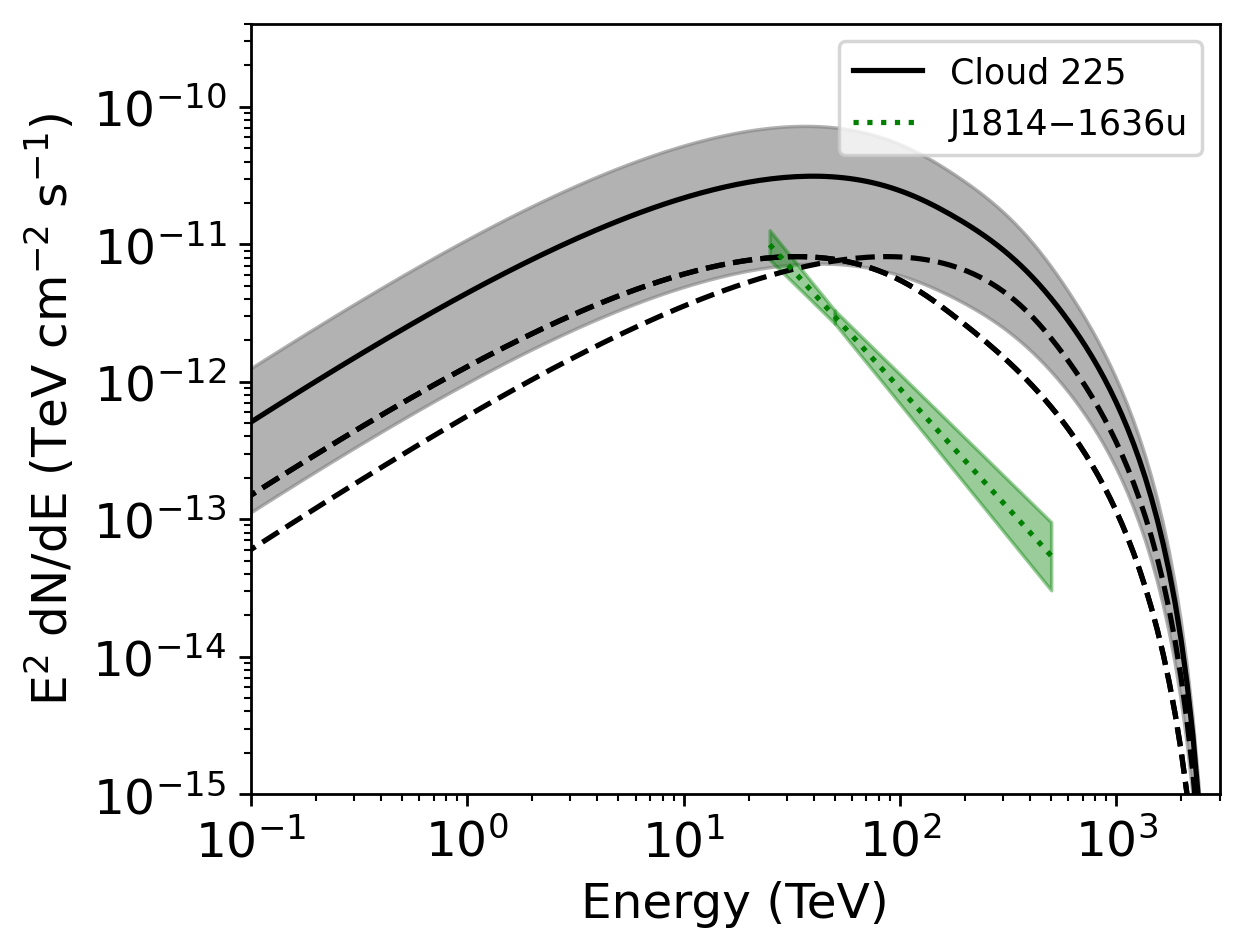}
    \put(25,68){\small type Ia}
    \end{overpic}
    \begin{overpic}[width=0.9\columnwidth]{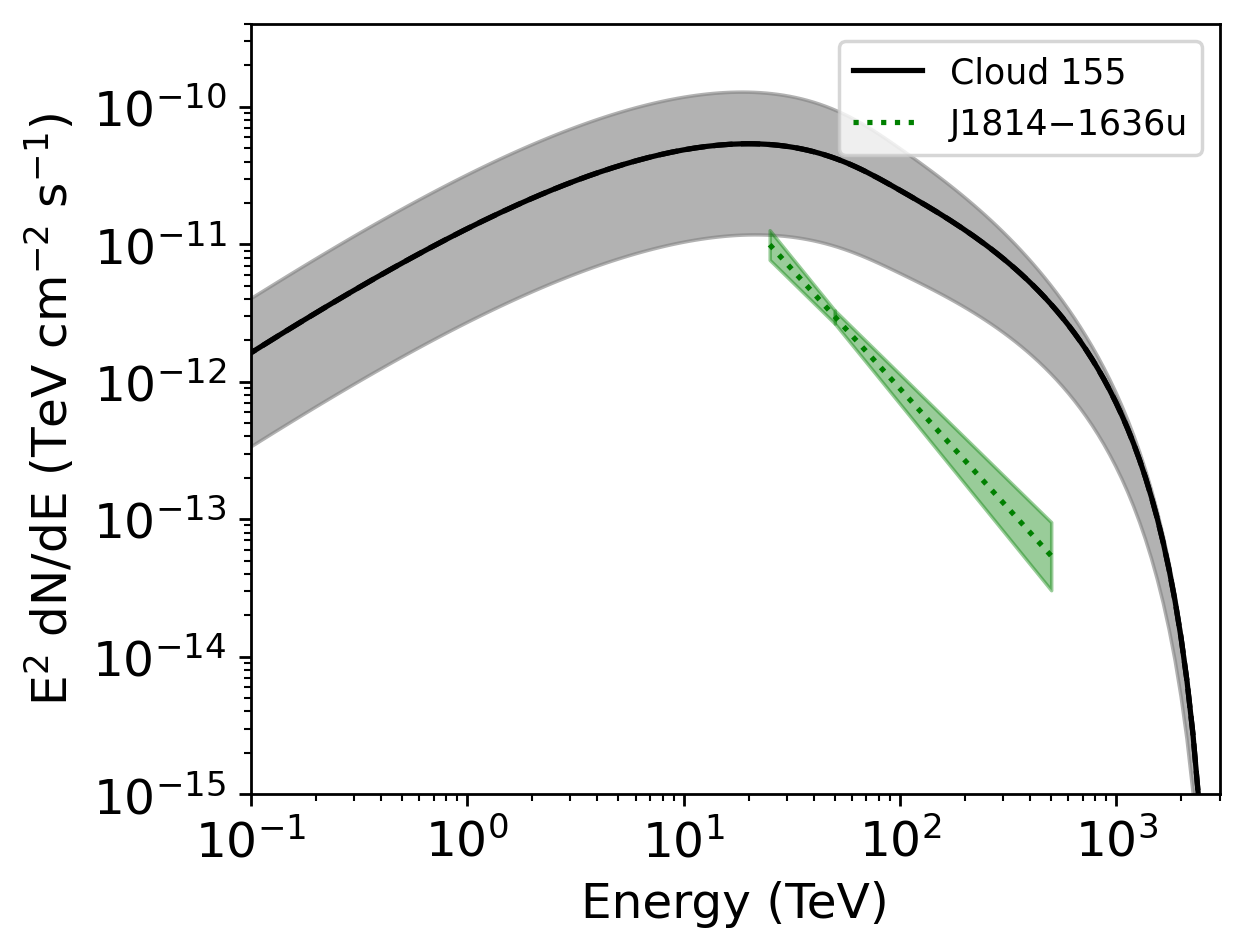}
    \put(25,68){\small type Ia}
    \end{overpic}
    \begin{overpic}[width=0.9\columnwidth]{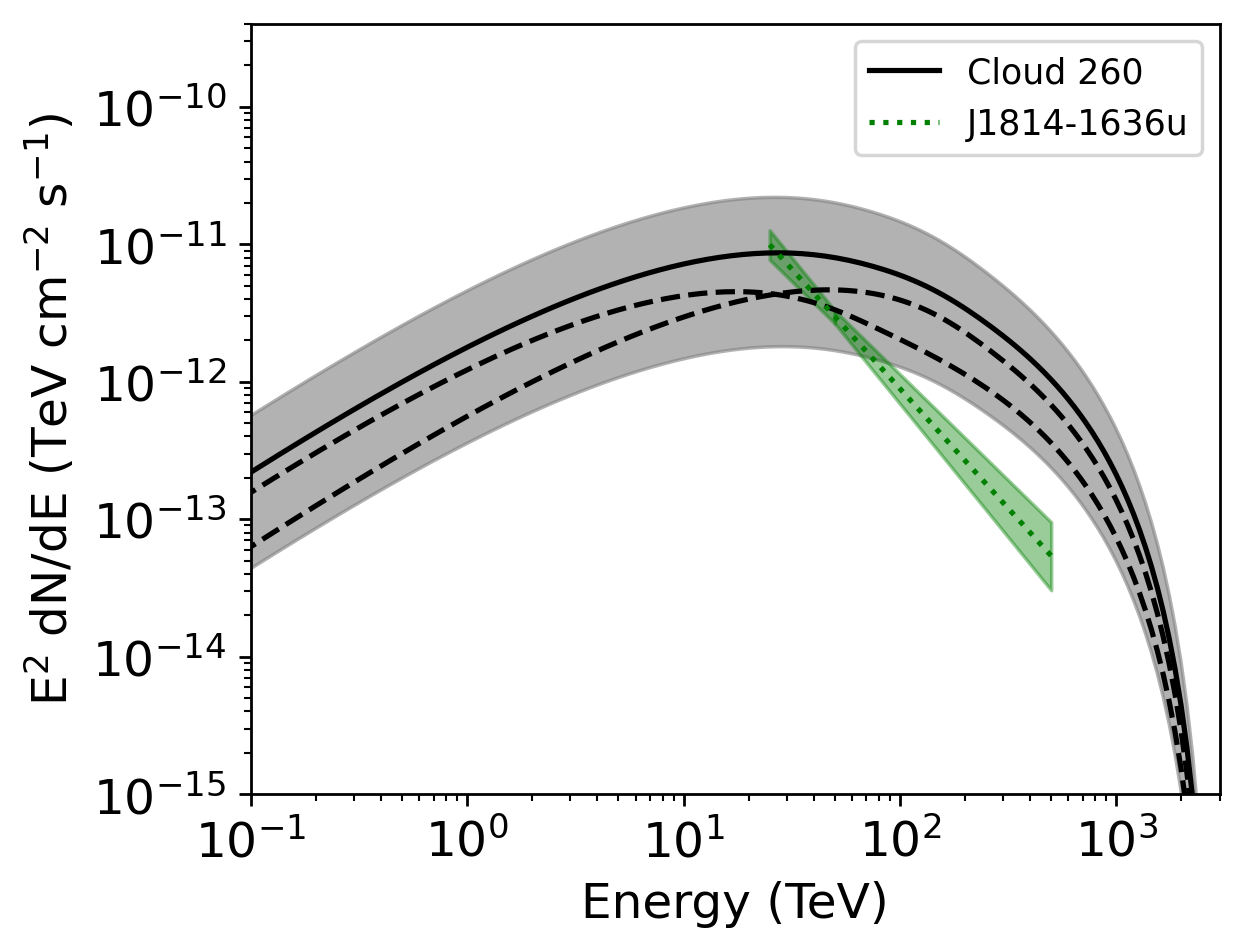}
    \put(25,68){\small type Ia}
    \end{overpic}
    \begin{overpic}[width=0.9\columnwidth]{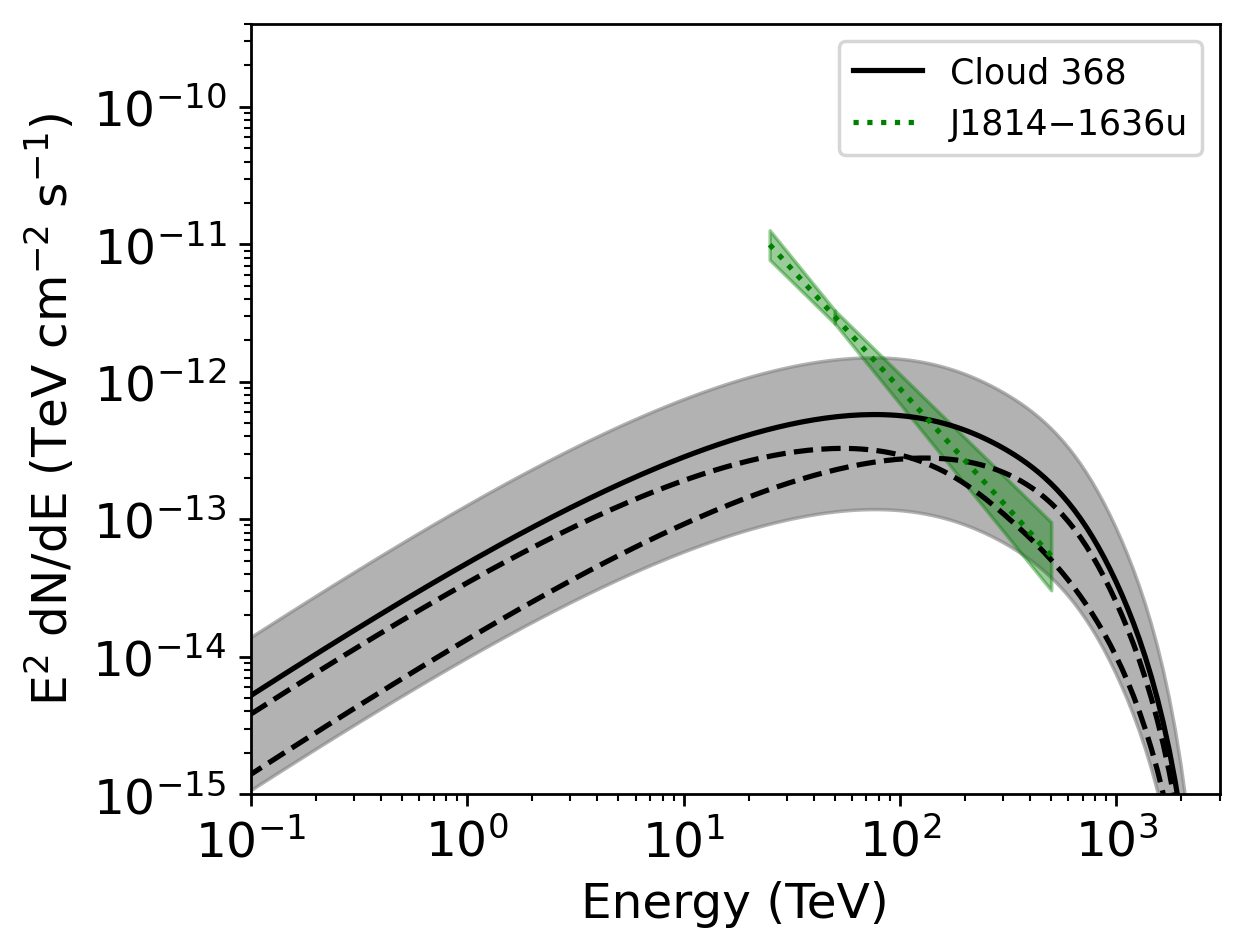}
    \put(25,68){\small type II}
    \end{overpic}
    \caption{As in Figure \ref{fig:c368_j1814} for 1LHAASO\,J1814-1636u, with the addition of cloud 260 at $(l,b)=(14.61^\circ,-0.34^\circ)$. }
    \label{fig:c201_j1814}
\end{figure*}

Figure \ref{fig:c201_j1814} shows 1LHAASO\,J1814-1636u compared to several clouds that are illuminated only under the type Ia SNe scenario. In both cases, multiple SNRs contribute to the total flux, for which no distance estimates are available. Although multiple sources may contribute towards the total gamma-ray emission along the line-of-sight, it is unlikely that all SNRs will be located at the same distance suitable to illuminate the respective cloud simultaneously. Cloud\,201 is situated at a distance of $\sim$3.59\,kpc, whilst cloud\,225 is located at $\sim$4.1\,kpc, i.e. for the same illuminating SNR (G012.7-0.0) we are assuming different distances in different plots. Hence the curves due to individual SNRs (dashed lines in figure \ref{fig:c201_j1814}) should also be considered, and have comparable uncertainties to that of the total emission.

\subsection{1LHAASO J1858+0330}

\begin{figure}
    \centering
    \begin{overpic}[width=0.9\columnwidth]{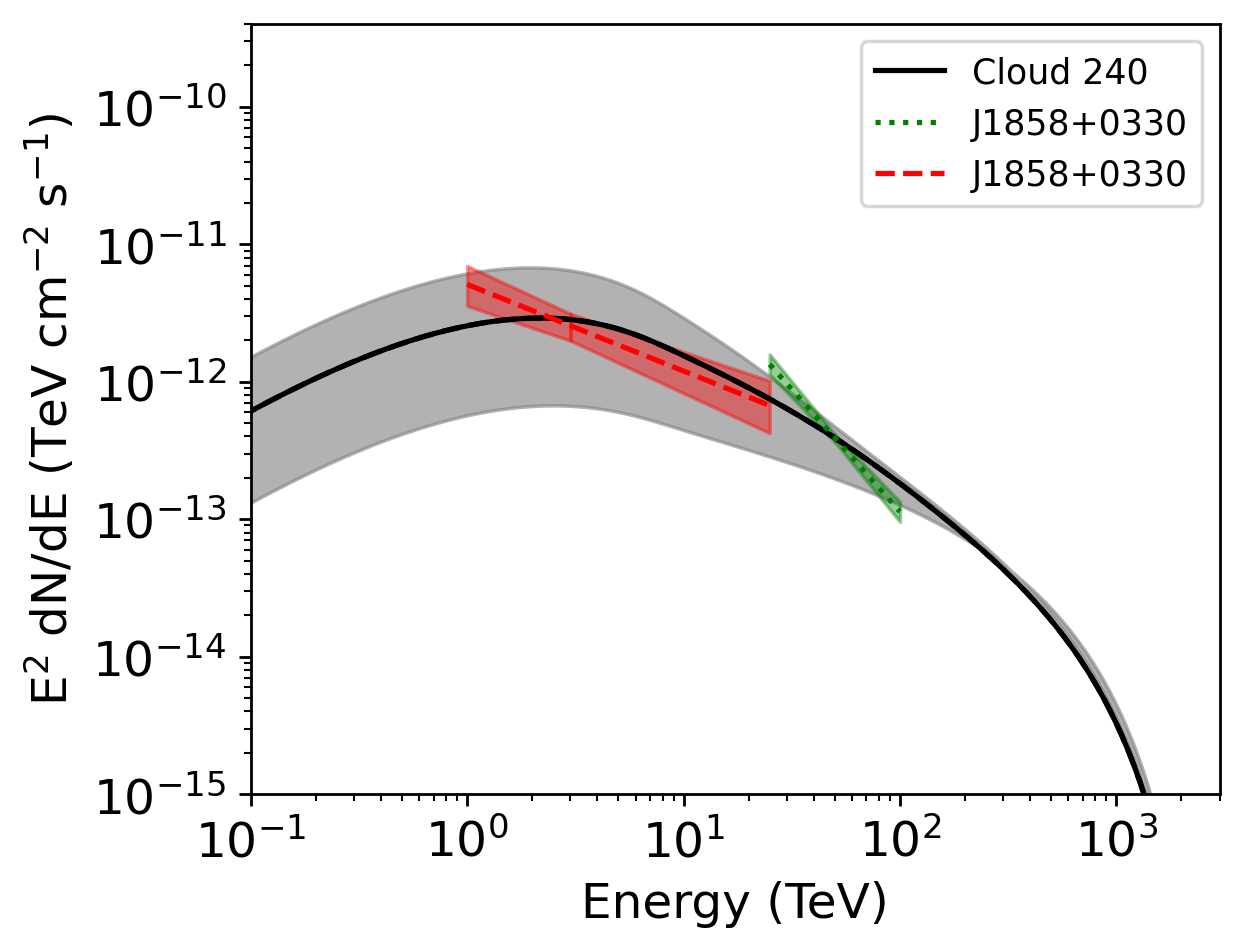}
    \put(25,68){\small type II}
    \end{overpic}

    \caption{As in figure \ref{fig:c190_j1857} for 1LHAASO\,J1858+0330 with cloud\,240, located at $(l,b)=(36.1^\circ,-0.14^\circ)$. }
    \label{fig:c240_j1858}
\end{figure}

Figure \ref{fig:c240_j1858} compares 1LHAASO\,J1858+0330 to cloud\,240, showing good agreement with the spectra measured by both WCDA and KM2A instruments of LHAASO. Although there are counterparts listed in \citet{2024ApJS..271...25Cao_1lhaaso} with Fermi-LAT sources, the source is unidentified with no reasonable physical associations known. Despite the good spectral match, we note here that the angular separation between the cloud and  1LHAASO\,J1858+0330 is somewhat larger than for the other sources coincident with this cloud (1LHAASO\,J1857+0245 and 1LHAASO\,J1857+0203u), such that an association of cloud\,240 with this source is speculative only.

\subsection{1LHAASO J1825-1256u}

\begin{figure}
    \centering
    \begin{overpic}[width=0.9\columnwidth]{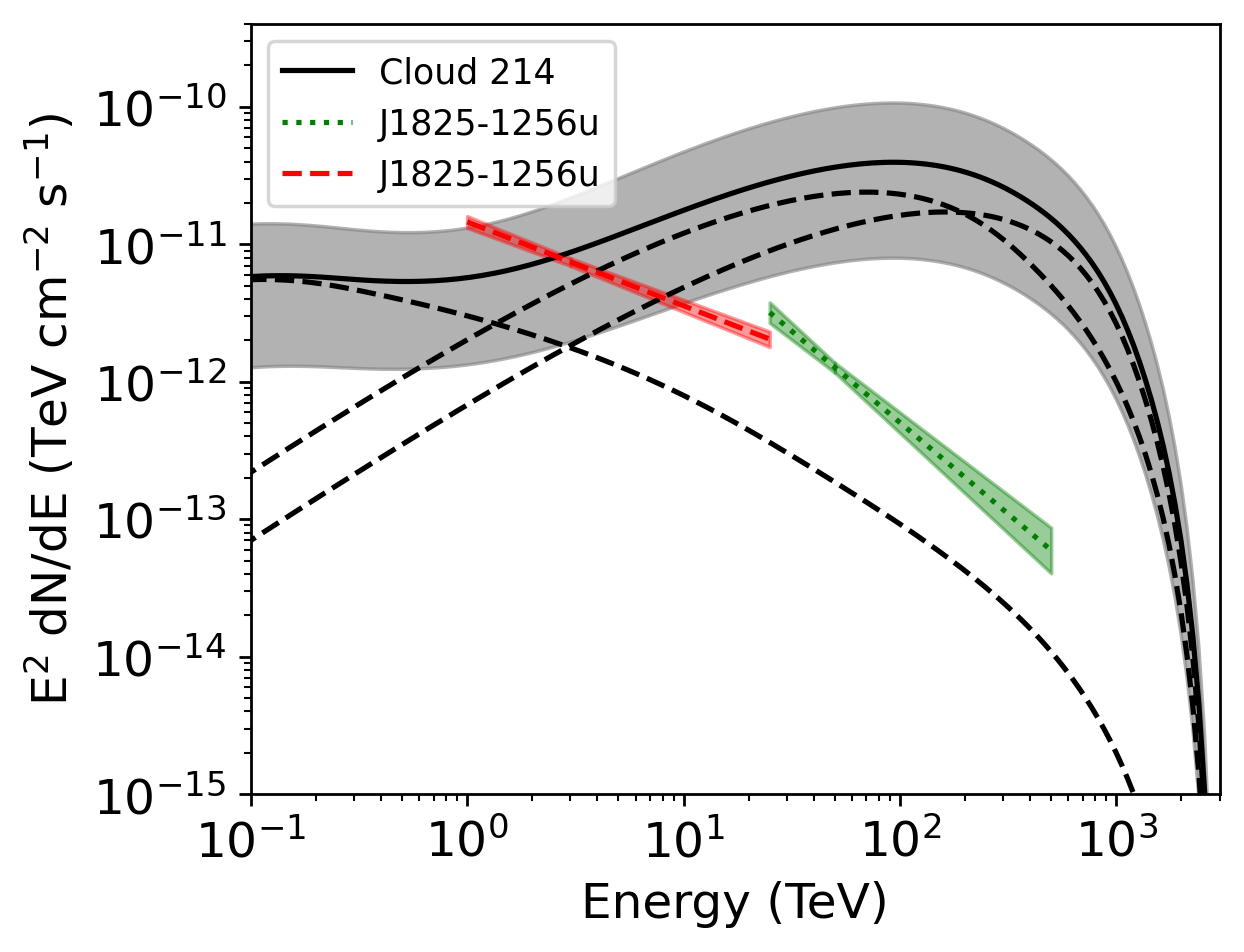}   
    \put(80,68){\small type Ia}
    \end{overpic}
    \caption{As in figure \ref{fig:c190_j1857} for 1LHAASO\,J1825-1256u with cloud\,214 located at $(l,b)=(18.4^\circ,-0.27^\circ)$. }
    \label{fig:c214_j1825}
\end{figure}

The region around 1LHAASO\,J1825-1256u is complex, with multiple bright TeV gamma-ray sources known. Whilst it is anticipated that known PWNe are responsible for the nearby 1LHAASO\,J1825-1418 and 1LHAASO\,J1825-1337u, the nature of 1LHAASO\,J1825-1256u remains obscure, either as part of the PWN emission or potentially an independent source. Indeed, the presence of an emission component coincident with a molecular cloud in the region has already been proposed by HAWC \citep{2021ApJ...907L..30A_J1825_HAWC}. The spatial coincidence of cloud 214 with 1LHAASO\,J1825-1256u is good, with two SNRs providing the bulk of the CR flux, namely G017.0-0.0 and G017.4-0.1. Little is known about either SNR, however whilst modelling of the former appears to severely overestimate the flux, modelling of the latter SNR provides a reasonable match to the spectral shape, albeit at a lower normalisation. This scenario is hence worth exploring in more detail for further studies of this complex sky region.

\subsection{1LHAASO sources with PWN associations}

Spectral energy distributions of the remaining 1LHAASO sources, typically addressed in connection with PWNe or halos (i.e. the environments of energetic pulsars), and model predictions for coincident clouds can be found in the appendix \ref{sec:sed_pwne}. In these cases, we consider a physical association with an SNR-illuminated cloud to be unlikely; nevertheless, even if the bulk of the gamma-ray emission is provided by inverse Compton scattering of leptons in the PWN, nearby clouds could still be illuminated by a CR flux originating from the progenitor SNR and contribute at the highest energies, as was for instance discussed in the case of the Crab nebula \citep{2022ApJ...924...42NieCrab}. 

\section{Conclusions and outlook}
\label{sec:conclude}

UHE gamma-ray astronomy has considerably boosted the search for PeVatrons, with SNRs, PWNe and superbubbles proving to be effective CR accelerators. Nevertheless, at UHEs, a new population of passive sources may emerge, due to interstellar clouds being illuminated by a flux of high-energy escaping CRs from nearby accelerators. Such a physical model was found to well describe the unidentified source LHAASO\,J2108+5157 in \cite{2024A&A...684A..66Mitchell}. We have shown this to be a viable scenario for producing gamma-ray emission that may contribute to the observed flux from several complex sky regions, yielding predictions for hadronic gamma-ray fluxes comparable to measurements by LHAASO from coincident sources. 

For many cases in figures \ref{fig:c190_j1857} to \ref{fig:c201_j1814}, we note that our model over-predicts the gamma-ray flux from the coincident 1LHAASO source. This demonstrates that one or more of our model assumptions are likely incorrect, such as the CR conversion efficiency, which we take to be 10\%. As this translates into the normalisation of the flux, an over-prediction implies that this strong assumption on the CR efficiency can be relaxed, although several parameters of the model can affect the normalisation.
Further significant uncertainties include the distance estimate of the SNR (which we take to be the distance of the cloud if it is unknown) and our assumptions on the particle transport in the ISM, which neglects spatial variation of the diffusion coefficient. It is anticipated that noticeable suppression of the diffusion coefficient occurs in the surroundings of the accelerator as well as in the clouds, hence we incorporated a suppression by up to a factor 100 in the uncertainties of the model. 

A major limitation of our approach is the necessary assumption of distance to the source, as the distance to LHAASO sources can only be constrained via physical associations and is not known apriori. Where multiple SNRs are assumed to have the same distance as a cloud, the expected emission due to each SNR individually should also be considered. 
However, we emphasise once again that these spectral energy distributions serve as a comparison only, whilst dedicated modelling for each source, required to refine and adapt the model to each case, is left for future studies. Given the level of agreement without adapting the model, the most promising candidate appears to be cloud\,240 with 1LHAASO\,J1857+0203u (figure \ref{fig:c190_j1857}). 
Our study is also limited by the available data on molecular target material in the corresponding sky regions, and more detailed follow-up studies, in particular concerning the properties of interstellar clouds, will be required to both improve the accuracy of our model and identify suitable counterparts that may explain the nature of the unidentified UHE sources.  \\

\textit{Acknowledgements}
AM is supported by the Deutsche Forschungsgemeinschaft, DFG project number 452934793. SC gratefully acknowledges financial support from Sapienza University with grant  ID RG12117A87956C66.

\bibliographystyle{elsarticle-harv} 
\bibliography{lhaaso_refs}

\newpage

\appendix

\section{Spatial overlap with PWN or pulsar halo sources}
\label{sec:sed_pwne}
\begin{figure*}
    \centering
    \begin{overpic}[width=0.9\columnwidth]{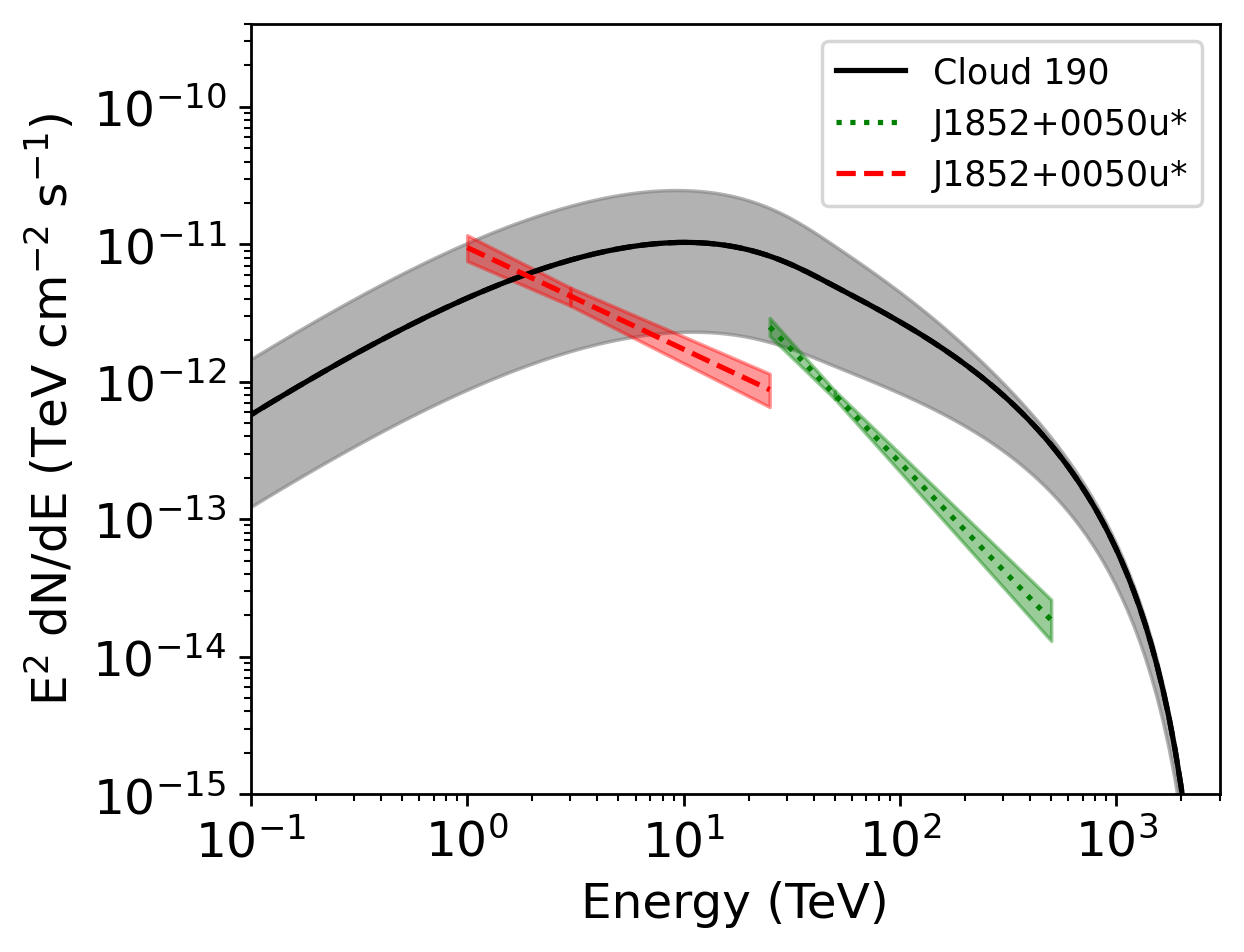}
    \put(25,68){\small type II}
    \end{overpic}
    \begin{overpic}[width=0.9\columnwidth]{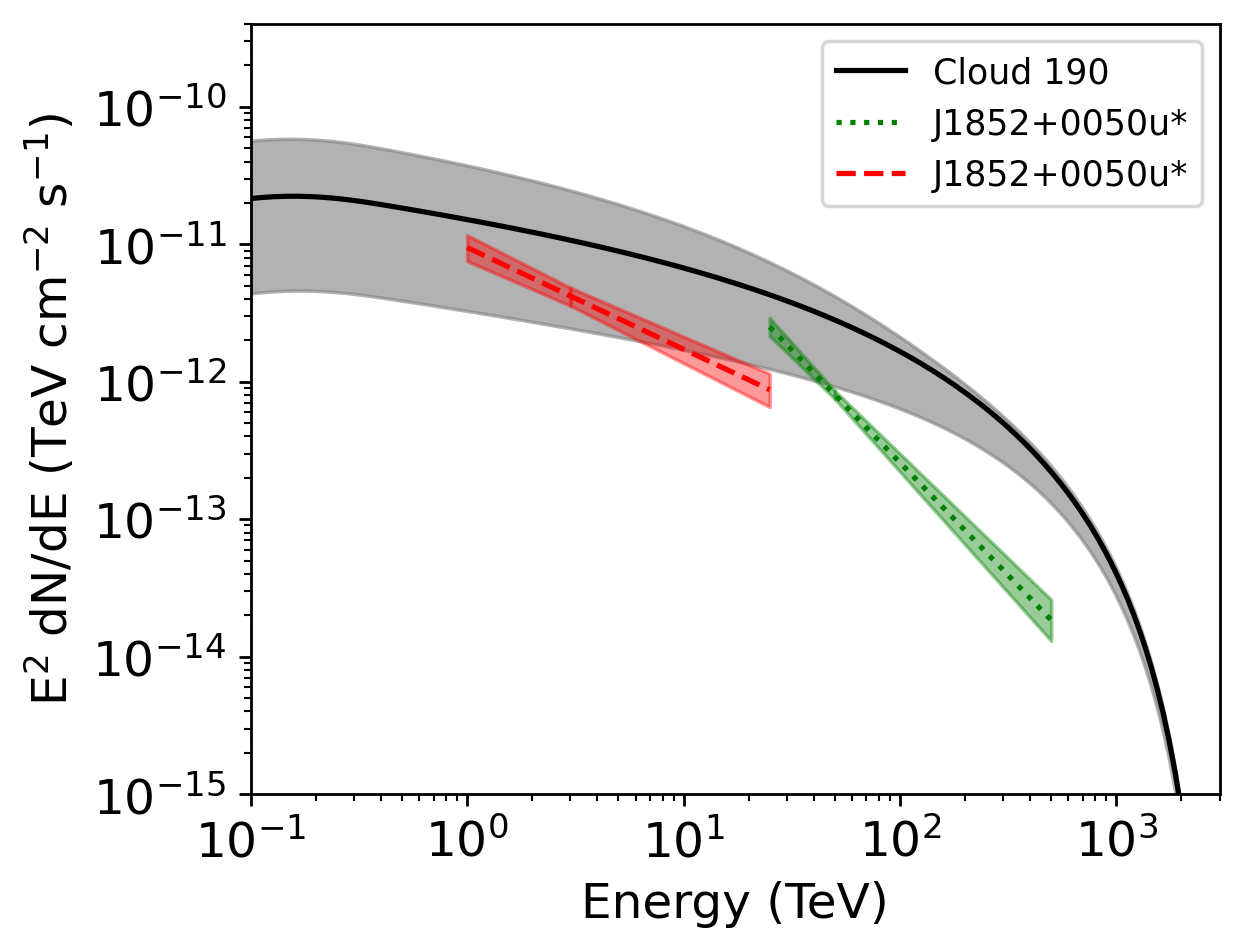}
    \put(25,68){\small type Ia}
    \end{overpic}
     \begin{overpic}[width=0.9\columnwidth]{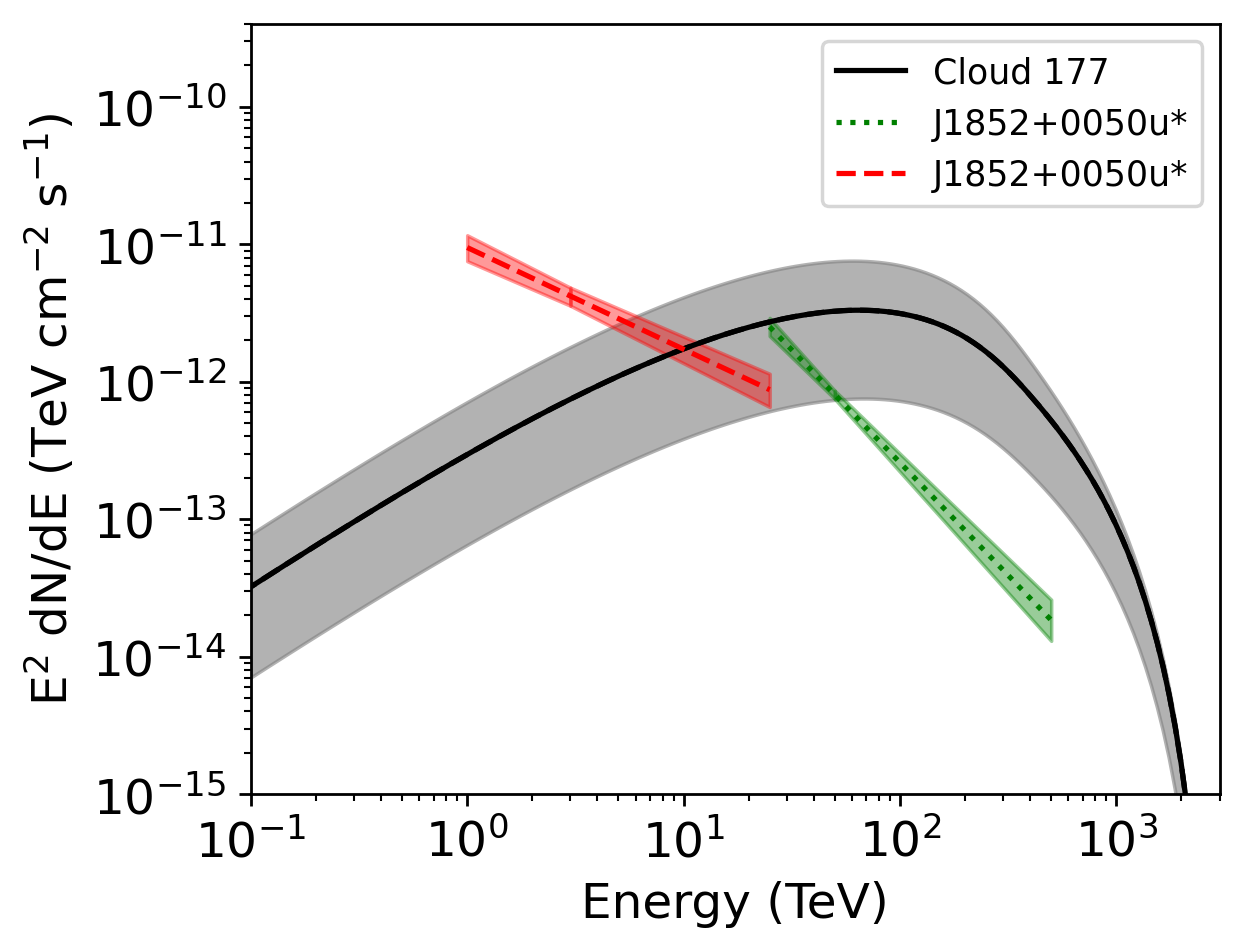}
     \put(25,68){\small type II}
    \end{overpic}
    \begin{overpic}[width=0.9\columnwidth]{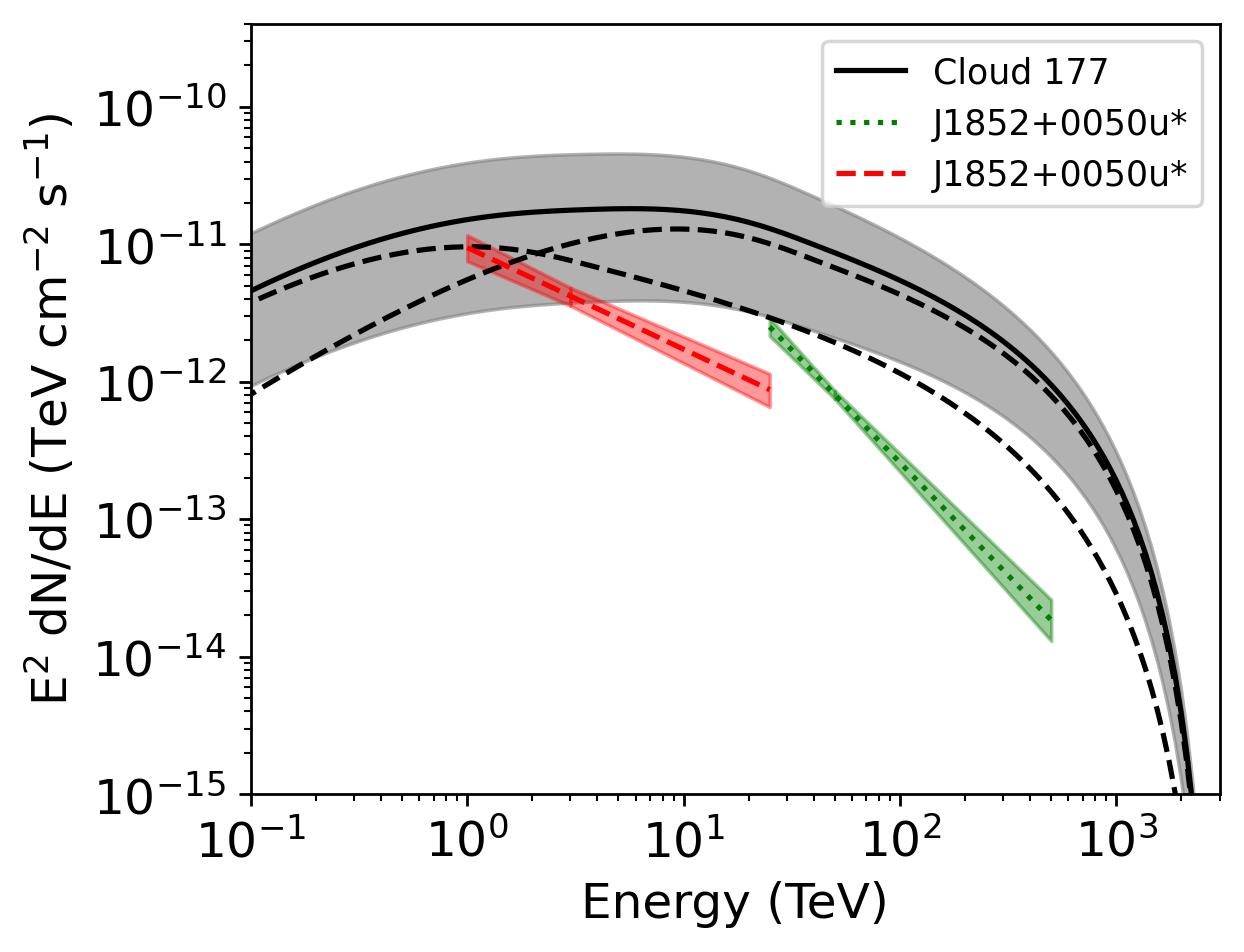}
    \put(25,68){\small type Ia}
    \end{overpic}
    \caption{As in figure \ref{fig:c190_j1857} for 1LHAASO\,J1852+0050u with clouds\,190 and 177 at $(l,b)=(34.99^\circ,-0.96^\circ)$, and $(l,b)=(35.64^\circ,0.01^\circ)$. } 
    \label{fig:c190_j1852}
\end{figure*}

\begin{figure*}
    \centering
    \begin{overpic}[width=0.9\columnwidth]{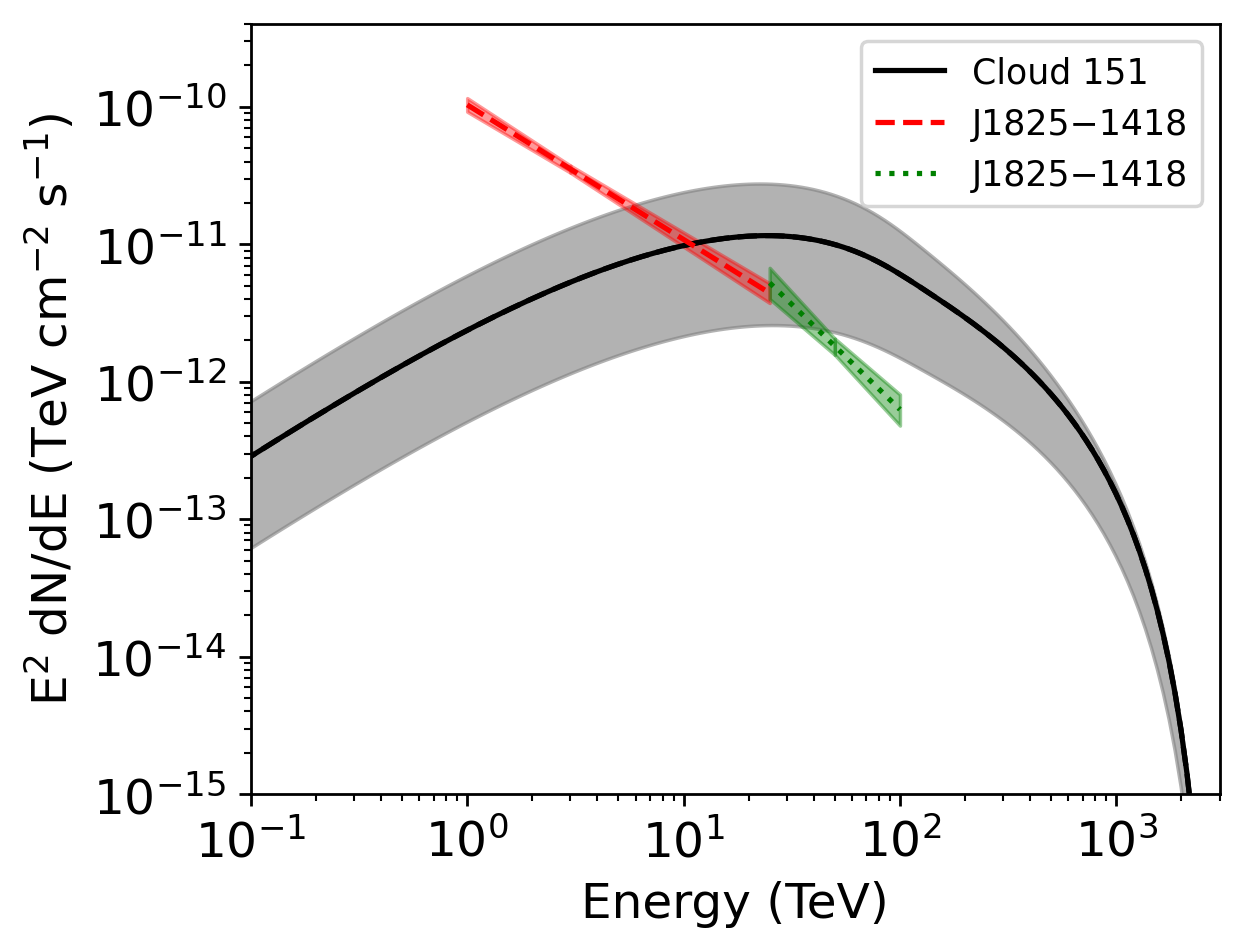}
    \put(25,68){\small type II}
    \end{overpic}
    \begin{overpic}[width=0.9\columnwidth]{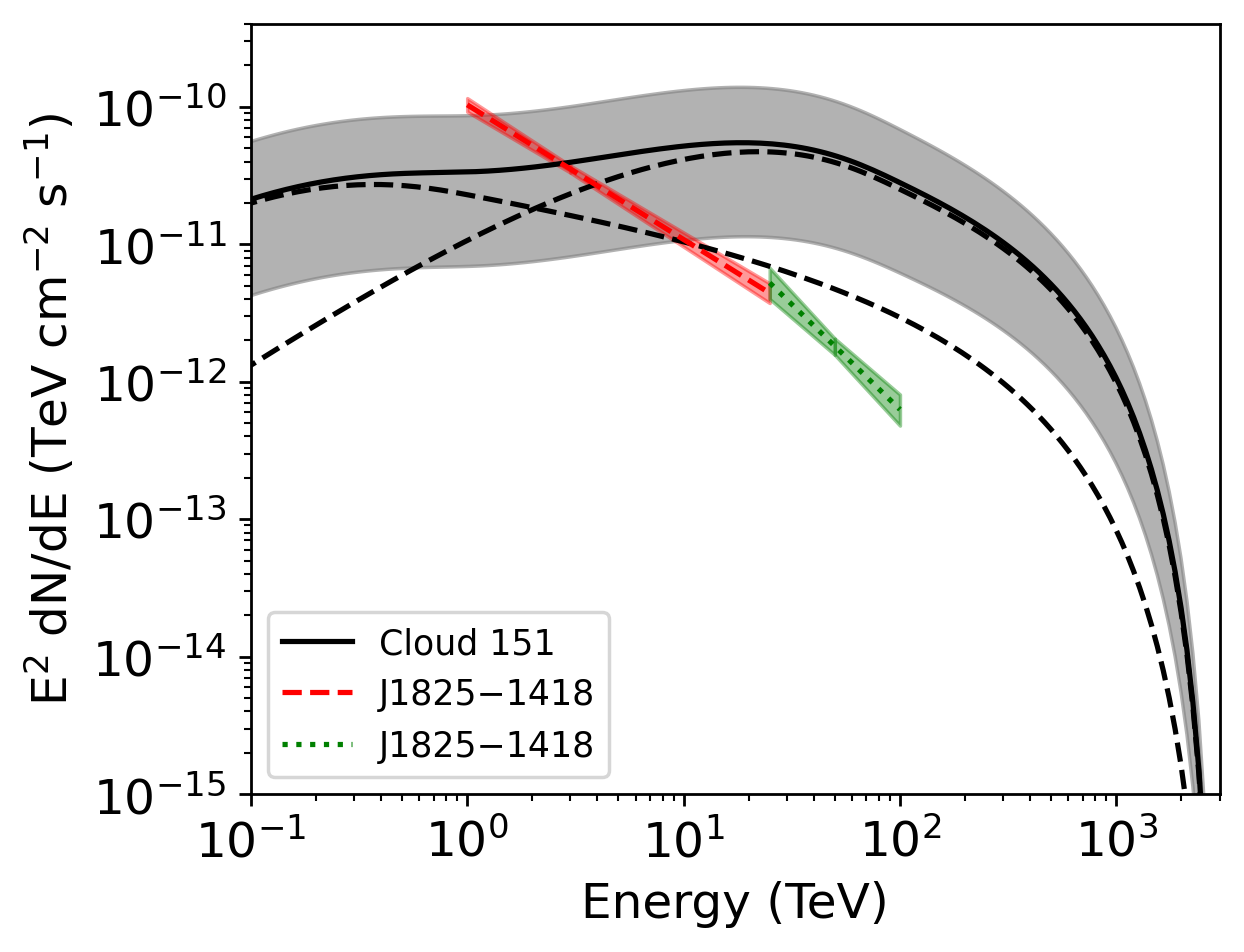}
    \put(25,68){\small type Ia}
    \end{overpic}
    \begin{overpic}[width=0.9\columnwidth]{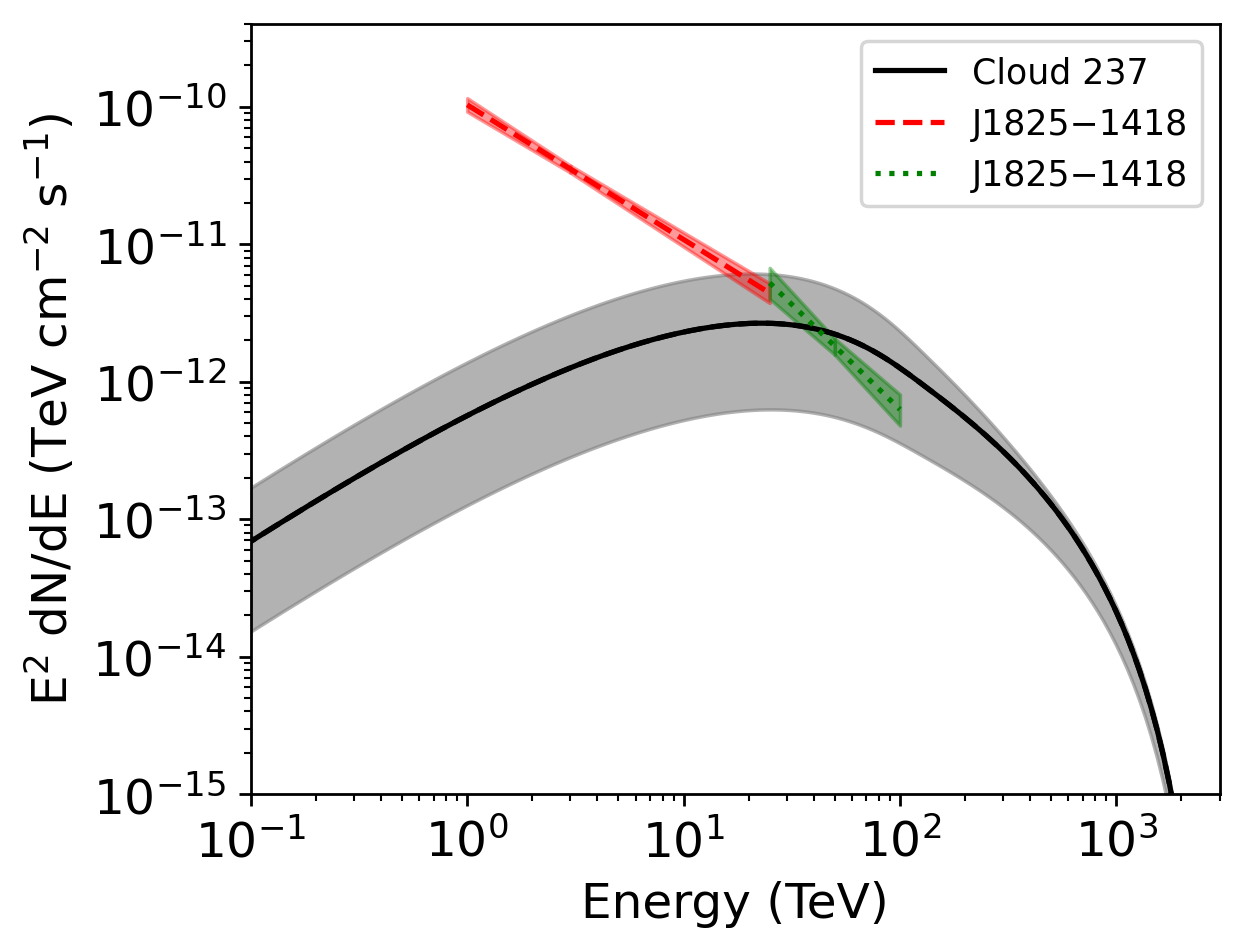}
    \put(25,68){\small type II}
    \end{overpic}
    \begin{overpic}[width=0.9\columnwidth]{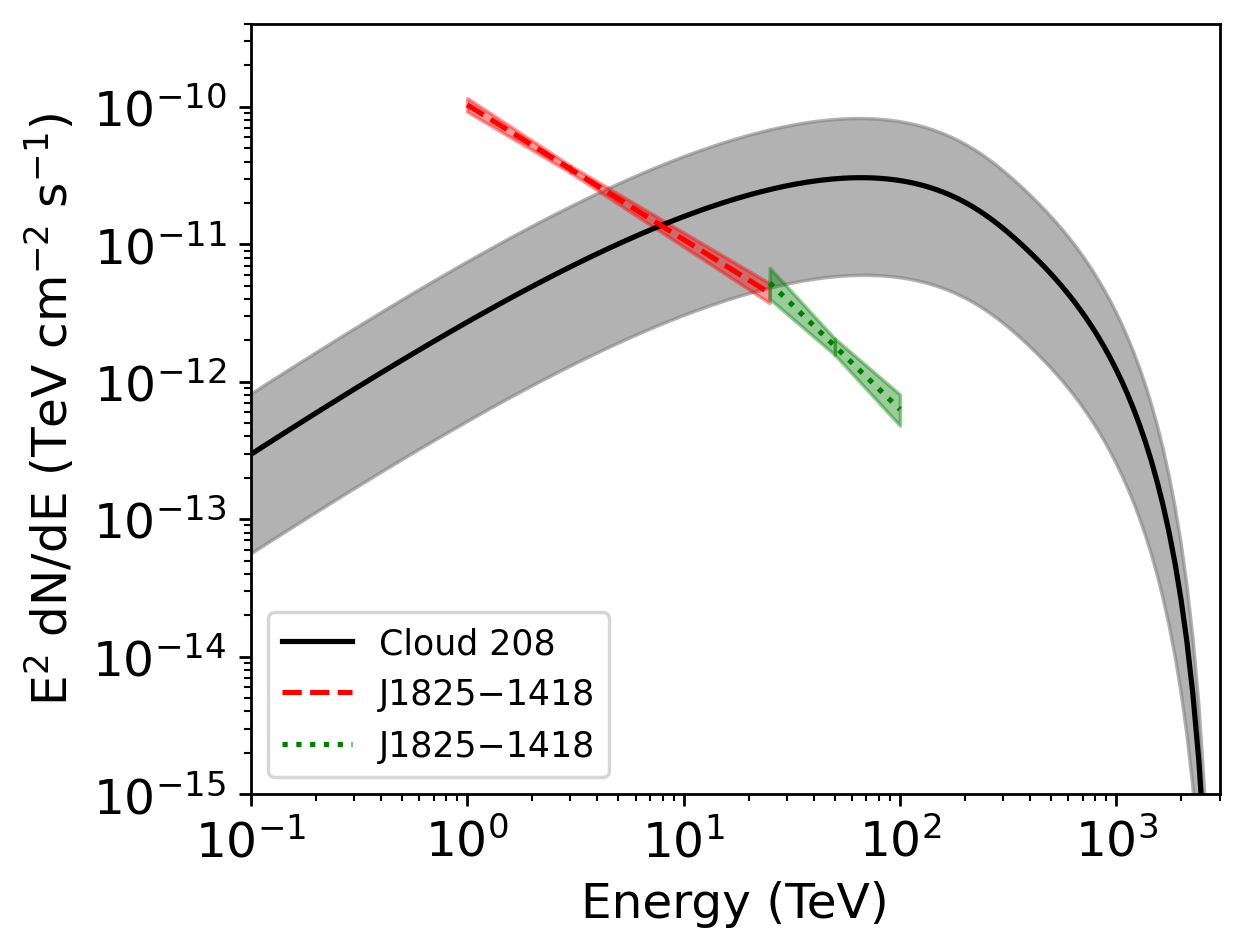}
    \put(25,68){\small type Ia}
    \end{overpic}
    \begin{overpic}[width=0.9\columnwidth]{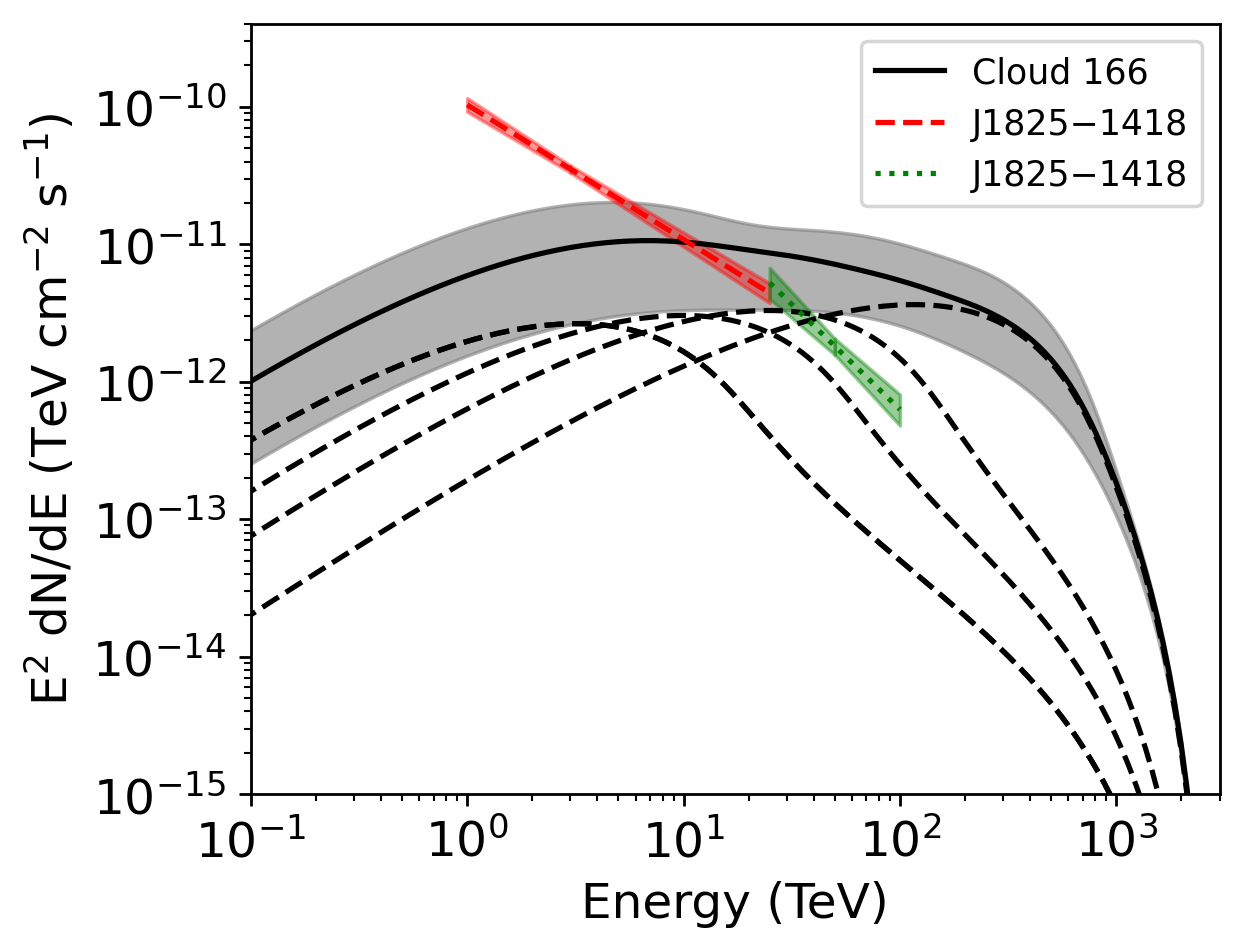}
    \put(25,68){\small type II}
    \end{overpic}
    \begin{overpic}[width=0.9\columnwidth]{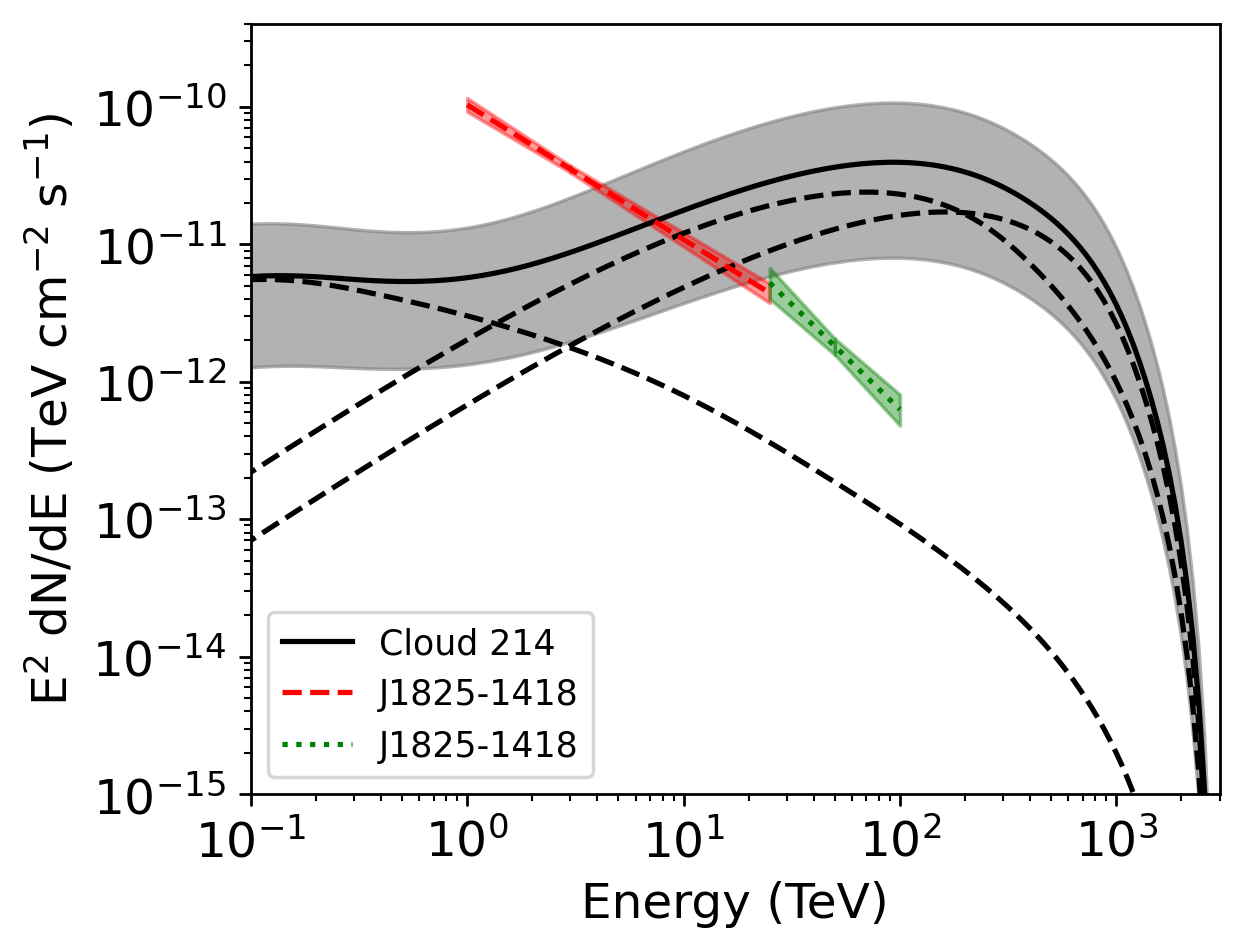}
    \put(25,68){\small type Ia}
    \end{overpic}
    
    \caption{As in figure \ref{fig:c190_j1857} for 1LHAASO\,J1825-1418, a large source with R$_{39}=0.81^\circ$ with clouds\,151 located at $(l,b)=(16.97^\circ,0.53^\circ)$, 237 at $(l,b)=(16.24^\circ,-1.02^\circ)$, 208 at $(l,b)=(16.61^\circ,-0.38^\circ)$, 166 at $(l,b)=(17.30^\circ,-1.40^\circ)$ and 214 at $(l,b)=(18.40^\circ,-0.27^\circ)$. \textit{Left:} type II SNe. \textit{Right:} type Ia SNe.}
    \label{fig:c151_j1825}
\end{figure*}

\begin{figure*}
    \centering
    \begin{overpic}[width=0.9\columnwidth]{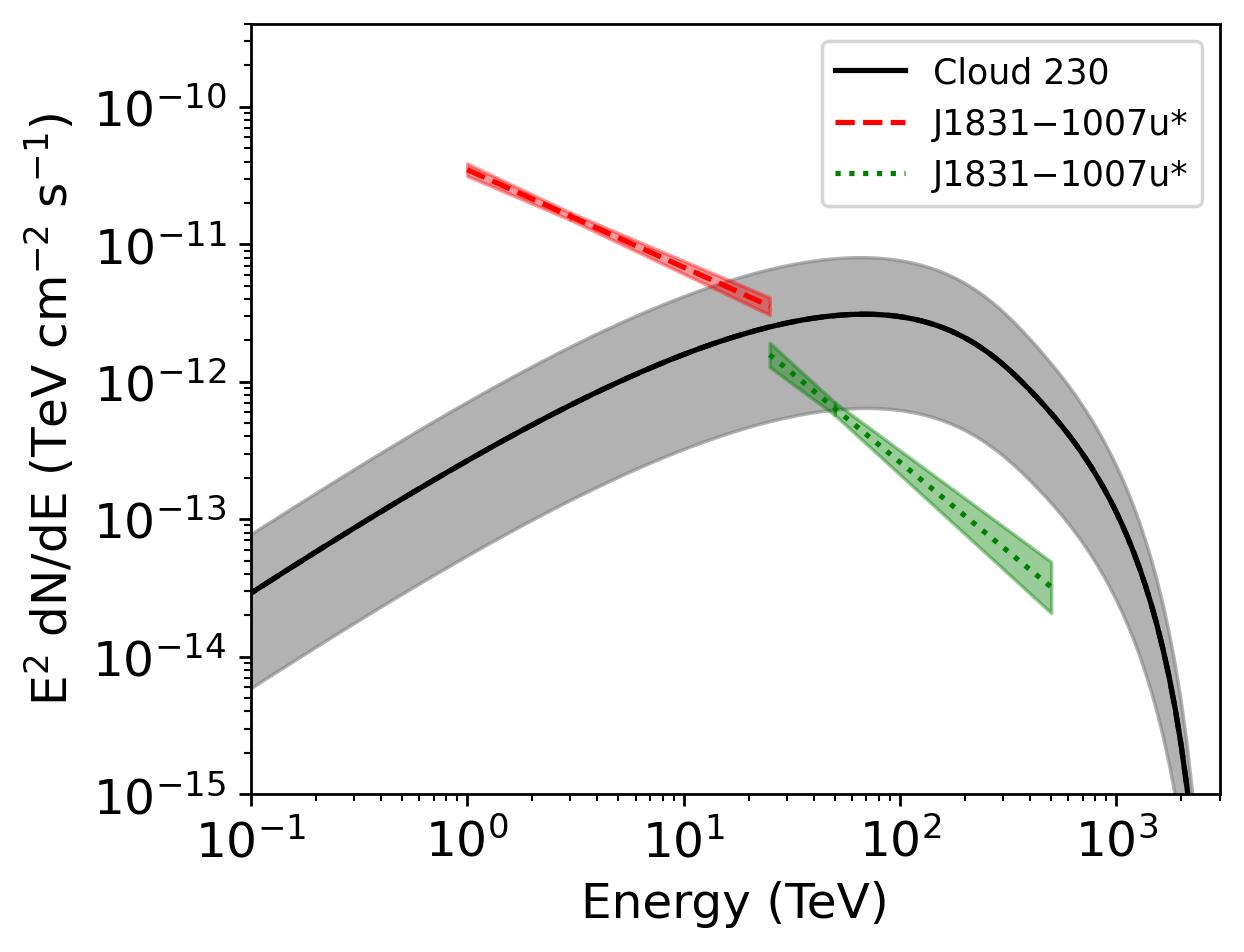}
    \put(25,68){\small type II}
    \end{overpic}
    \begin{overpic}[width=0.9\columnwidth]{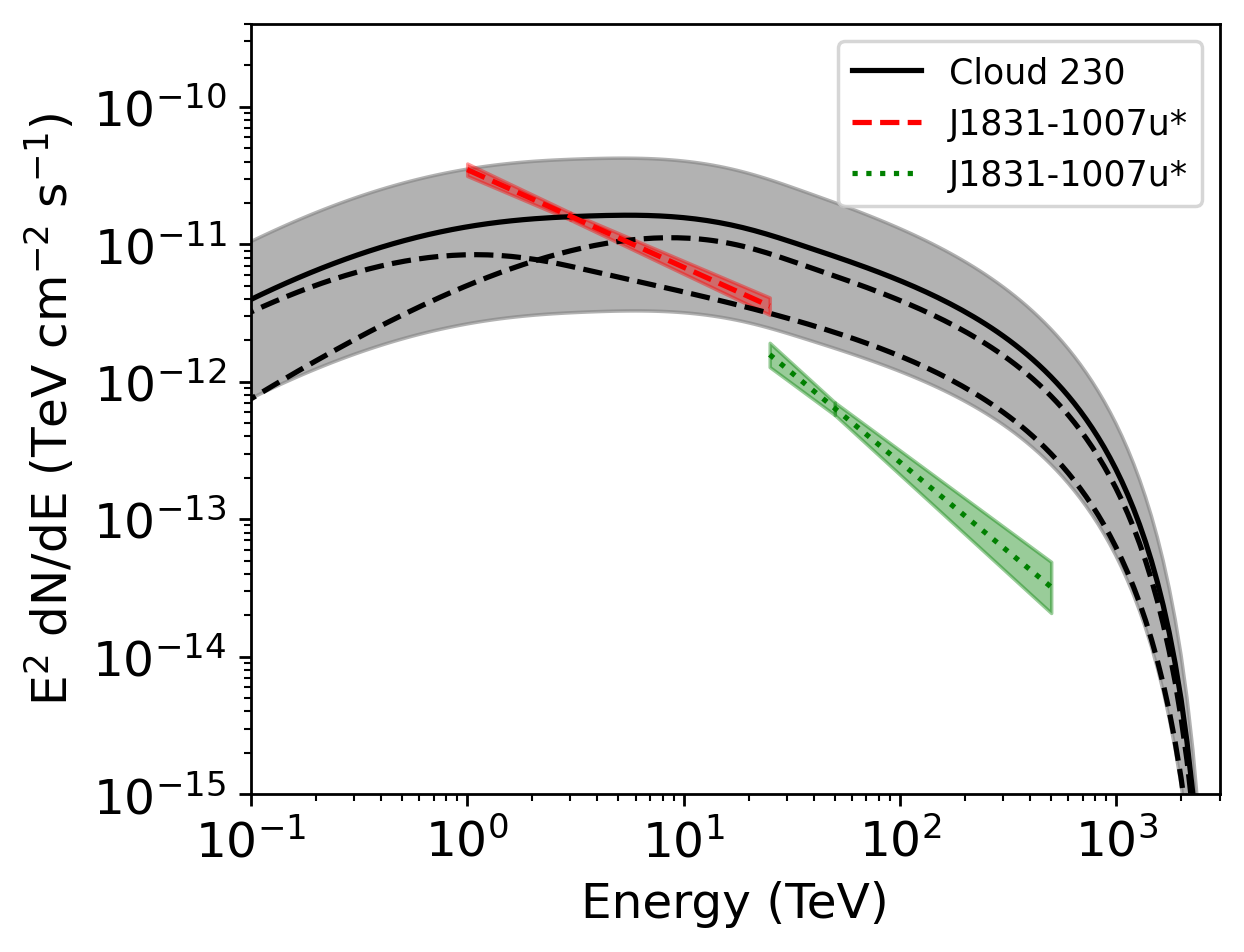}
    \put(25,68){\small type Ia}
    \end{overpic}
        \begin{overpic}[width=0.9\columnwidth]{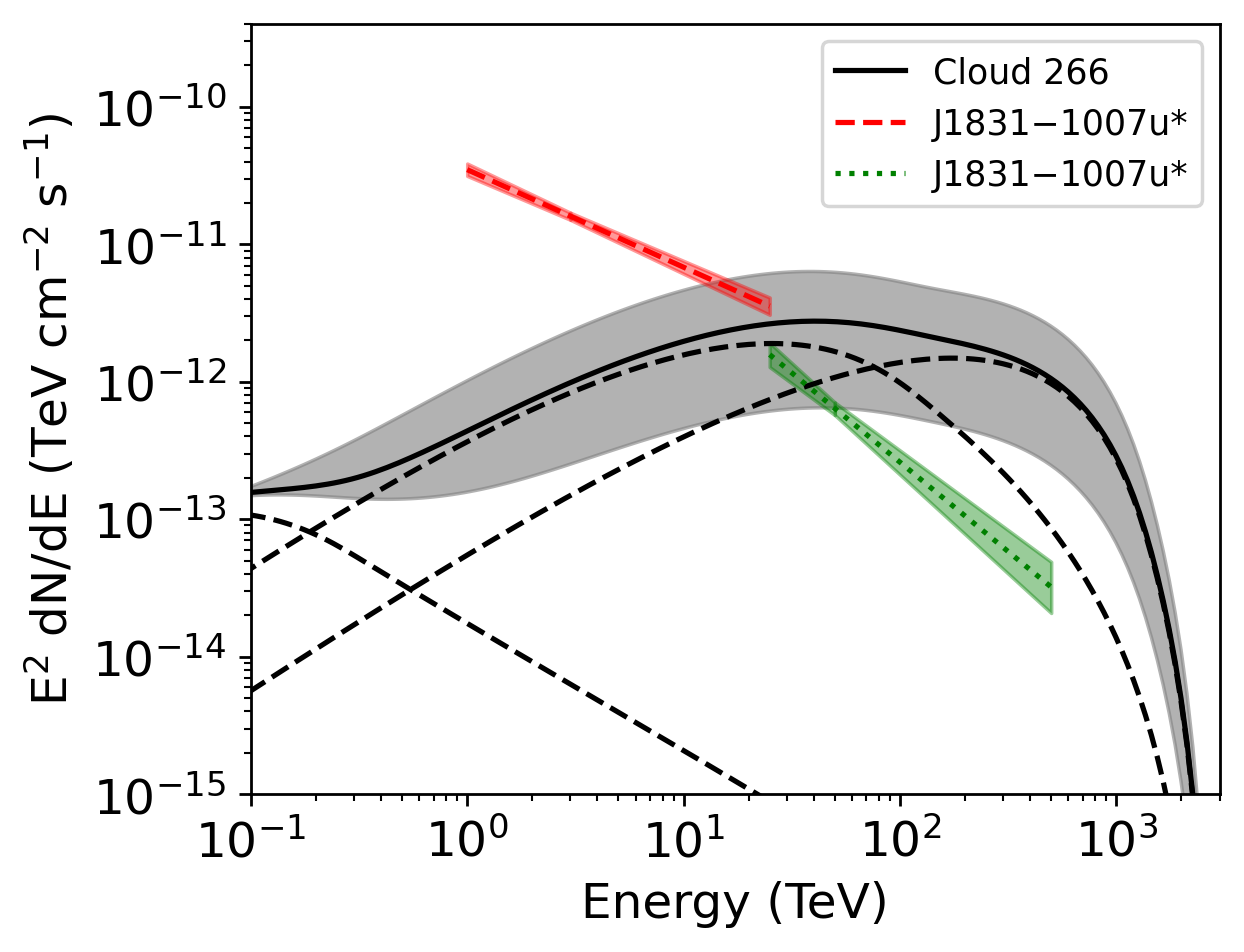}
    \put(25,68){\small type II}
    \end{overpic}
    \begin{overpic}[width=0.9\columnwidth]{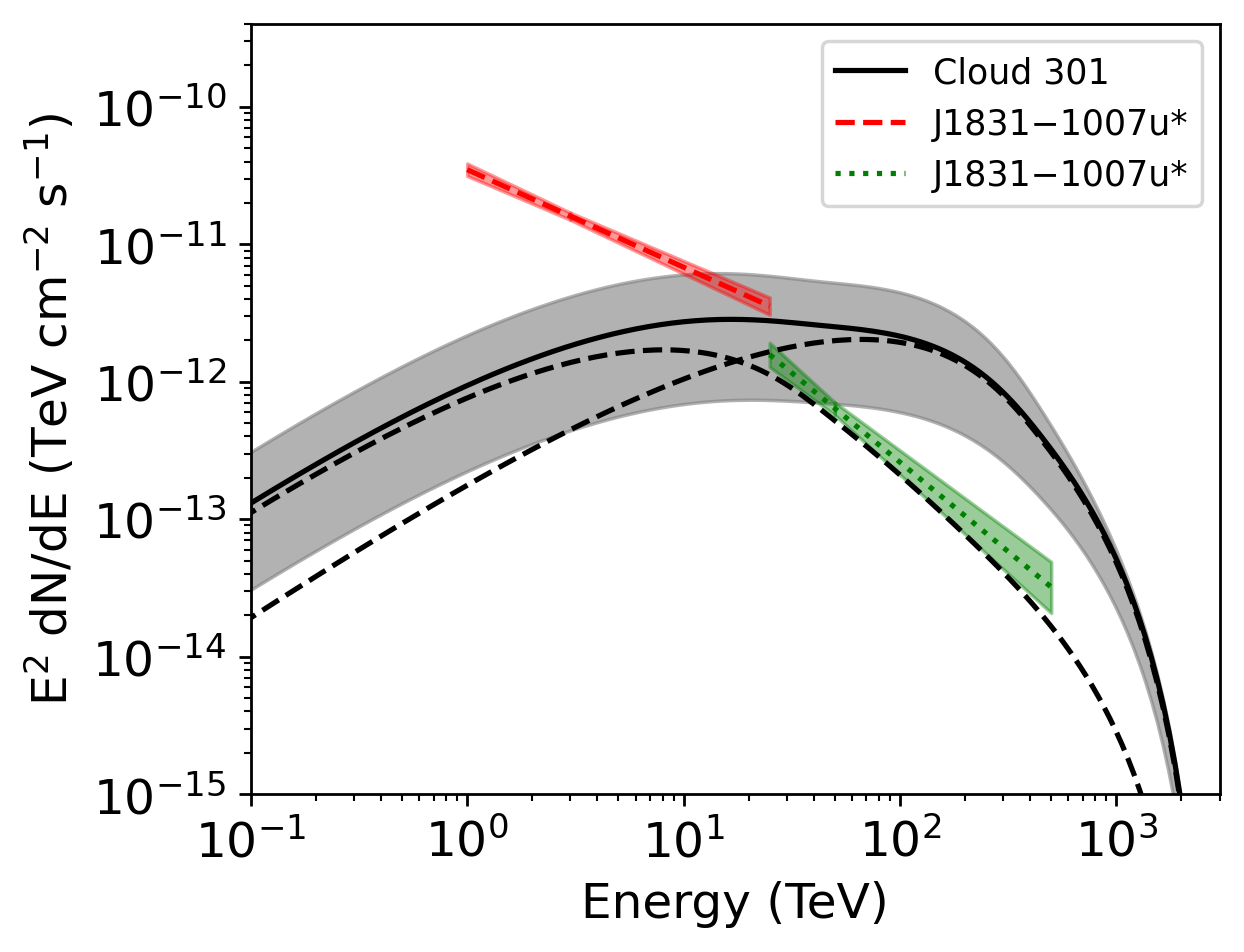}
    \put(25,68){\small type II}
    \end{overpic}
        \begin{overpic}[width=0.9\columnwidth]{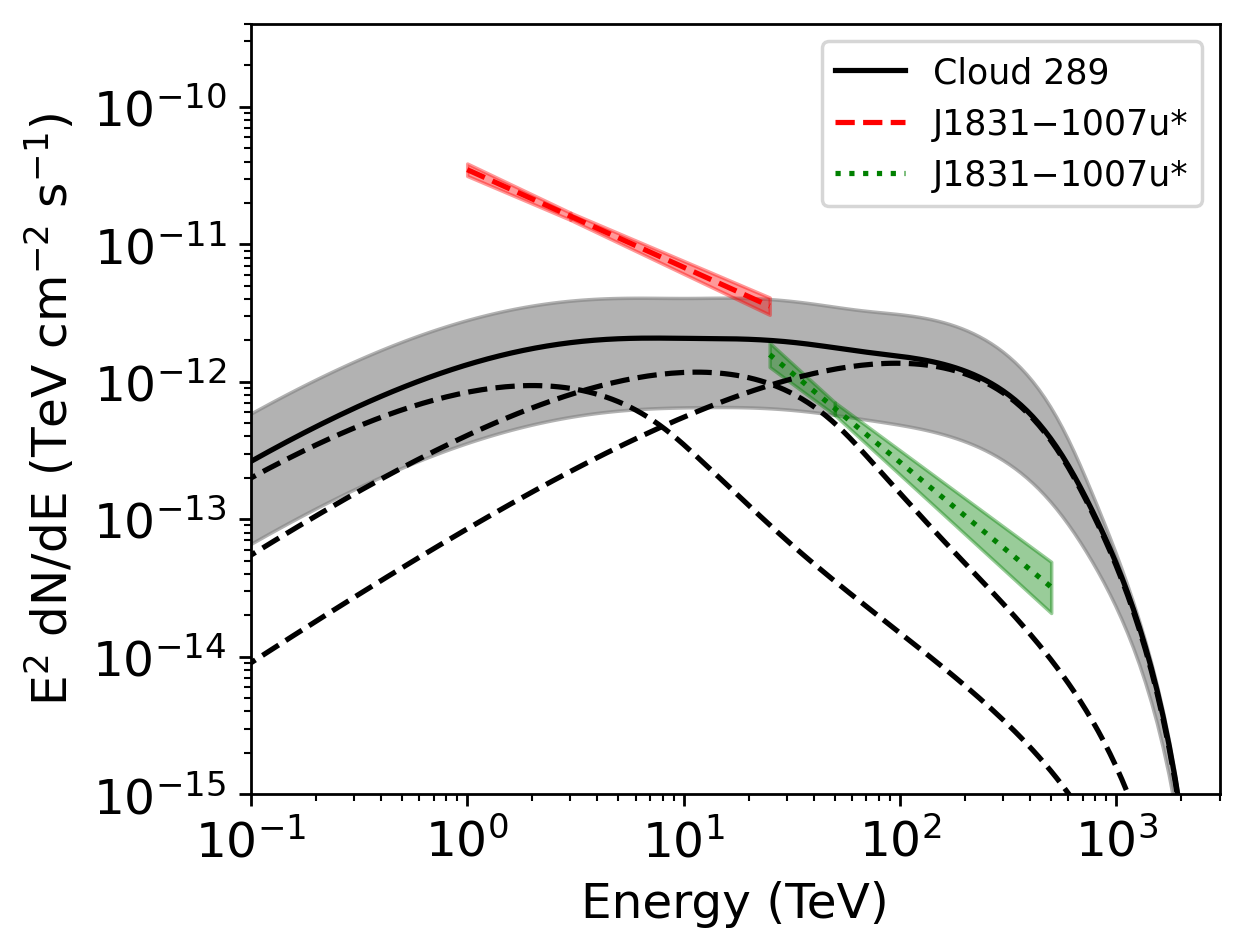}
    \put(25,68){\small type II}
    \end{overpic}
    \caption{As in figure \ref{fig:c190_j1857} for 1LHAASO\,J1831-1007u* with clouds\,230, located at $(l,b)=(21.97^\circ,-0.29^\circ)$, 266 at $(l,b)=(22.10^\circ,-1.06^\circ)$, 301 at $(l,b)=(21.78^\circ,-0.4^\circ)$ and 289 at $(l,b)=(21.25^\circ,0.02^\circ)$. }
    \label{fig:c230_j1831}
\end{figure*}

\begin{figure*}
    \centering
    \begin{overpic}[width=0.9\columnwidth]{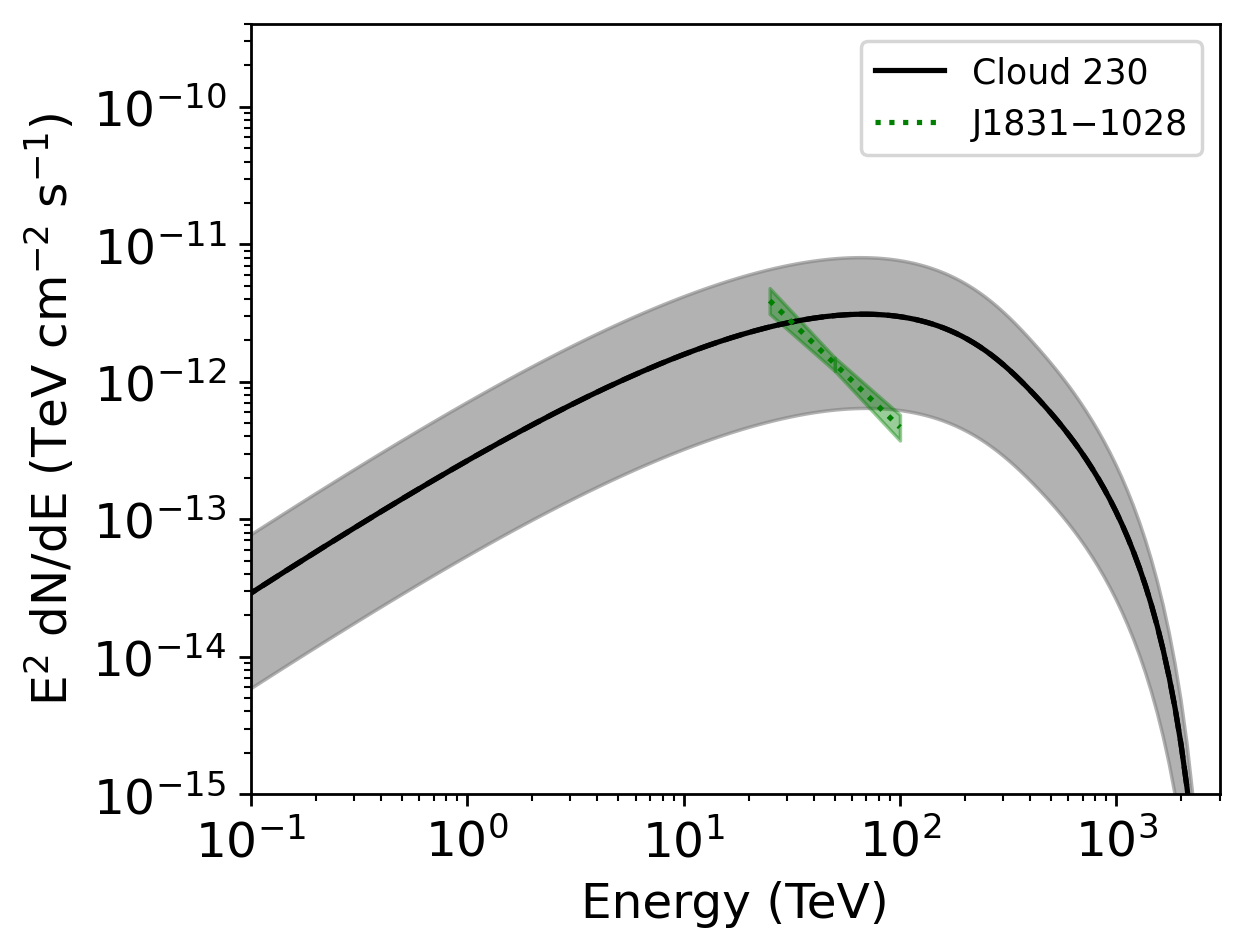}
    \put(25,68){\small type II}
    \end{overpic}
    \begin{overpic}[width=0.9\columnwidth]{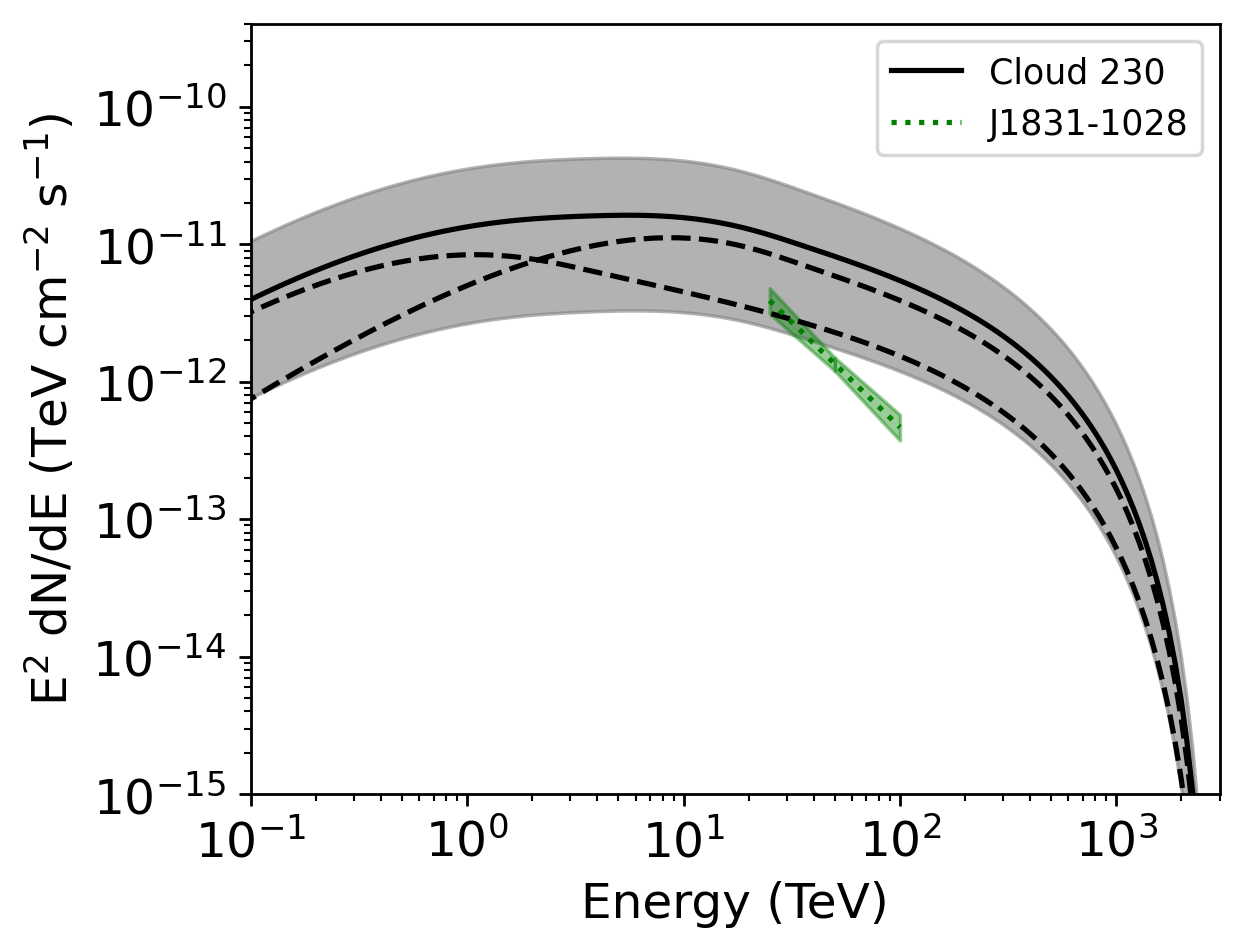}
    \put(25,68){\small type Ia}
    \end{overpic}
    \begin{overpic}[width=0.9\columnwidth]{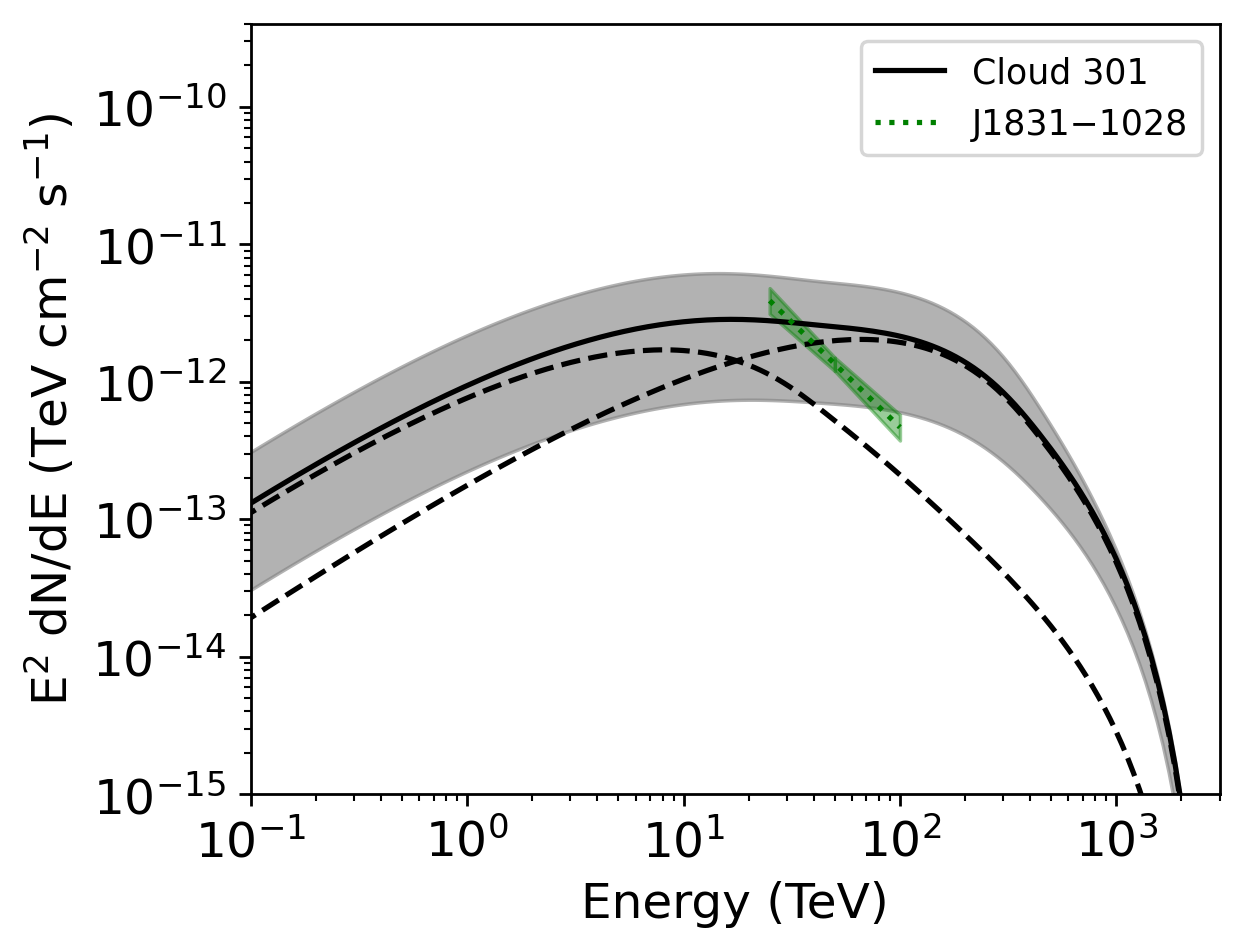}
    \put(25,68){\small type II}
    \end{overpic}
    \begin{overpic}[width=0.9\columnwidth]{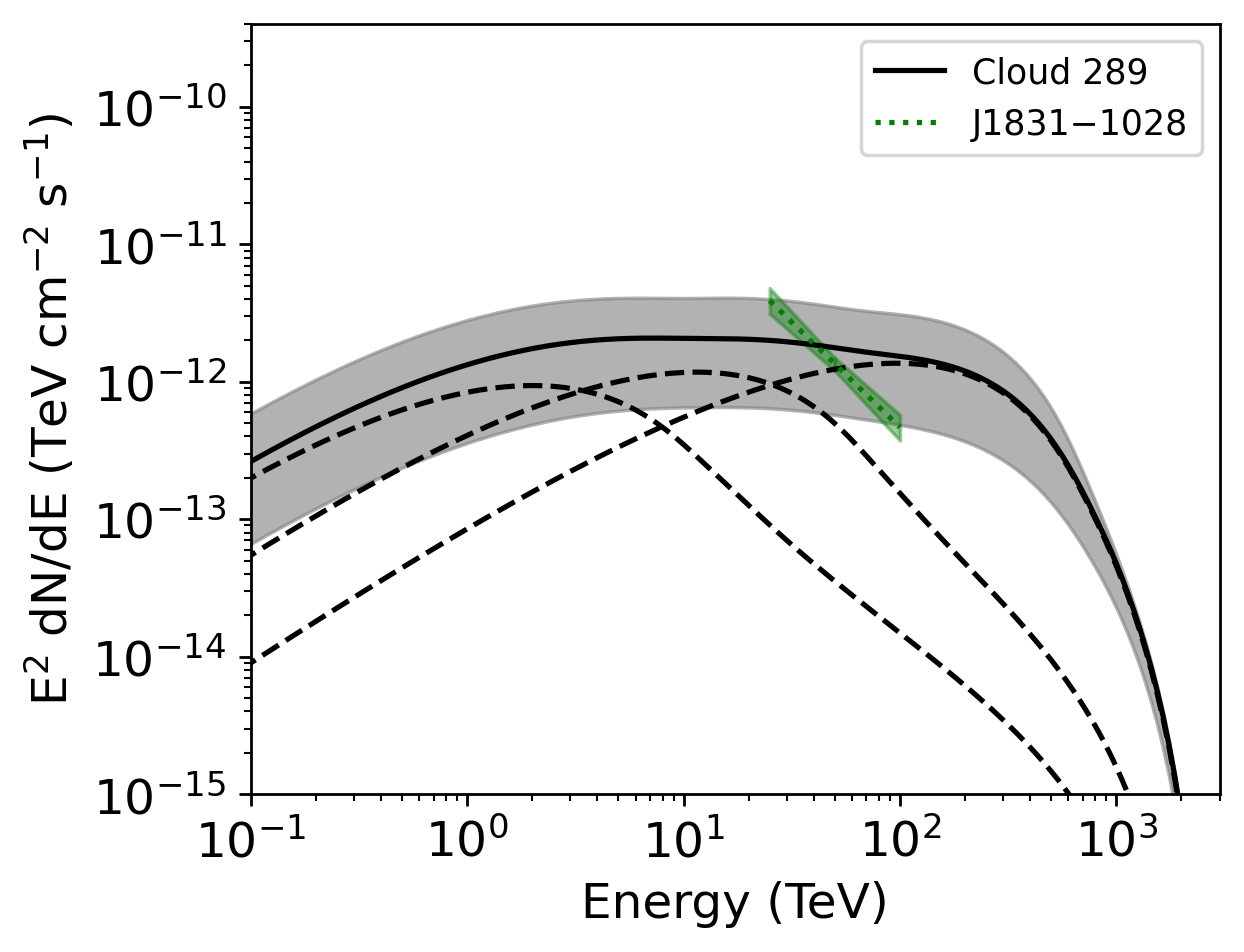}
    \put(25,68){\small type II}
    \end{overpic}

    \caption{As in figure \ref{fig:c230_j1831} for 1LHAASO\,J1831-1028 with clouds 230 at $(l,b)=(21.97^\circ,-0.29^\circ)$, 301 at $(l,b)=(21.78^\circ,-0.40^\circ)$ and 289 at $(l,b)=(21.25^\circ,0.02^\circ)$. }
    \label{fig:c301_j1831}
\end{figure*}

Figures \ref{fig:c190_j1852} to \ref{fig:c301_j1831} show the remaining clouds listed in tables \ref{tab:coincident} and \ref{tab:coincidentIA}, that are coincident with LHAASO sources putatively classified as PWNe or pulsar halos in \citep{2024ApJS..271...25Cao_1lhaaso}. As such, the illuminated cloud scenario is considered a less likely explanation for the origin of $\gamma$-ray emission in these sources.
As a result of this spectral comparison, there are no clear examples of compatibility between 1LHAASO sources associated with PWNe and illuminated clouds from known nearby SNRs.

\end{document}